\documentclass[conference]{IEEEtran}

\usepackage{graphicx}
\usepackage{algorithm}
\usepackage{amssymb}
\usepackage{amsmath}
\graphicspath{ {figures/} }

\begin{document}
\title{SCOR: Software-defined Constrained\\ Optimal Routing Platform for SDN }

\author{\IEEEauthorblockN{Siamak Layeghy\IEEEauthorrefmark{1},
Farzaneh Pakzad\IEEEauthorrefmark{2} and Marius Portmann\IEEEauthorrefmark{3}}
\IEEEauthorblockA{School of ITEE, 
The University of Queensland\\
Brisbane, Australia\\
Email: \IEEEauthorrefmark{1}siamak.layeghy@uq.net.au,
\IEEEauthorrefmark{2}farzaneh.pakzad@uq.net.au,
\IEEEauthorrefmark{3}marius@ieee.org
}}

\maketitle

\begin{abstract}
A Software-defined Constrained Optimal Routing (SCOR) platform is introduced as a Northbound interface in SDN architecture. It is based on constraint programming techniques and is implemented in MiniZinc modelling language. Using constraint programming techniques in this Northbound interface has created an efficient tool for implementing complex Quality of Service routing applications in a few lines of code. The code includes only the problem statement and the solution is found by a general solver program. A routing framework is introduced based on SDN's architecture model which uses SCOR as its Northbound interface and an upper layer of applications implemented in SCOR. Performance of a few implemented routing applications are evaluated in different network topologies, network sizes and various number of concurrent flows.
\end{abstract}


%
\IEEEpeerreviewmaketitle


\section{Introduction}\label{introduction}
Software Defined Networking (SDN) is introduced to improve network programmability, simplify network management and enhance network operation. It is a paradigm shift in networking that suggests  to decouple the control plane from the data plane and place the control of the system in a logically centralised node called `SDN controller'. Consequently, SDN controller has a view of the entire network resulting in the rapid deployment of the network changes \cite{mckeown2009software, feamster2013road, PRS-SDN_A_Comprehensive_Survey}.   

\begin{figure}[!t]
\centering 
\includegraphics [width=6cm, height=6cm, natwidth=400, natheight=545]{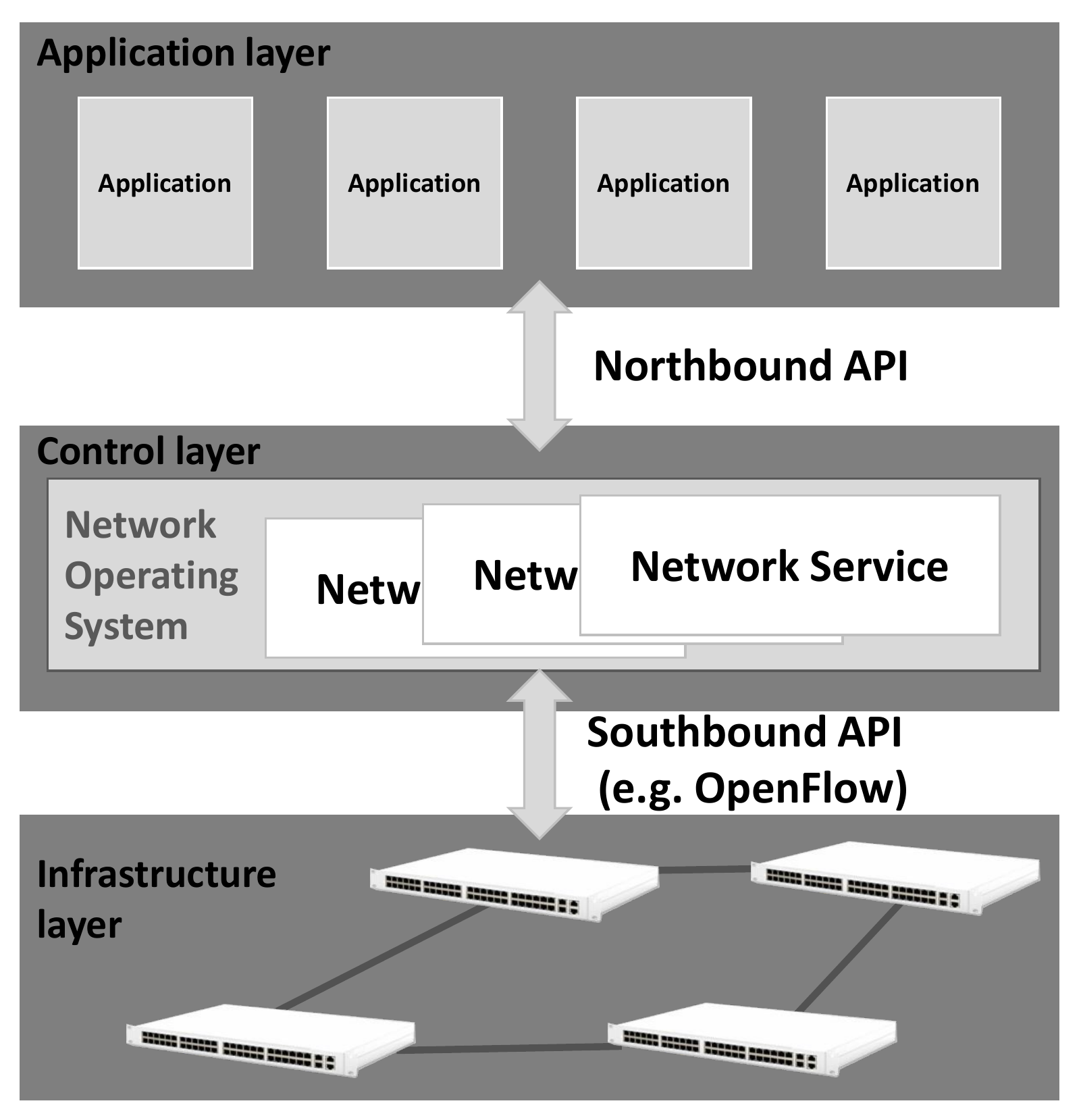}
\caption{SDN Architecture \cite{citeulike12475417}}
\label{fig:SDNarchitecture}
\end{figure}

Fig. \ref{fig:SDNarchitecture} shows the logical view of  SDN architecture. Data plane or forwarding elements, such as switches and routers, are located at the bottom layer. The next layer up is the control plane or Network Operating System (NOS) which has a whole view of the network and the ability to install, update and delete forwarding rules on forwarding elements. At the top layer is the application layer where the high level policy decisions such as routing, traffic engineering, and security are implemented  \cite{citeulike12475417}.

The control plane, SDN controller, communicates with the data plane via a Southbound interface, as shown in Fig. \ref{fig:SDNarchitecture}. The most dominant  Southbound interface is OpenFlow, which allows the SDN controller to install, update and remove flow rules on the flow tables of the switches. It also make it possible for the switches to send OpenFlow messages to the controller \cite{mckeown2009software, urlOpenFlow2014}. The communication between the controller and network applications is made possible using another interface, the Northbound Application Programming Interface (API). It enables applications to program the network and / or request services from it. 
There are a range of Northbound APIs which provide various functionalities though a standard Northbound interface is not introduced yet \cite{Jammal201474}. 

Several network applications are proposed in SDN which address different \emph{Traffic Engineering (TE)} requirements such as load balancing and link utilisation optimisation. Today's networking environment necessitates using various \emph{Quality of Service (QoS)} routing and TE techniques to optimise traffic for different media applications like voice and video. The implementation of these QoS routing and TE applications might be complex in some cases. Since there is no standard Northbound interface these applications require various implementation compatible with various SDN controllers. Despite the vital importance of QoS routing and TE in handling media traffic and network optimisation, none of the proposed Northbound interfaces provides specific abstractions or functions for implementing them.

In order to address this concern, this work is introducing a Software-defined Constrained Optimal Routing (SCOR) platform. This includes a northbound interface which is based on \emph{Constraint Programming (CP)} techniques that have been successfully applied in similar problems e.g. vehicle routing. The main idea behind using CP methods for this Northbound API is to separate complexity from the user. In CP, users do not specify step by step solution of the problem, as in the case of procedural programming. They only state the constraints that the solution should have. In other words, in CP, user states the problem and the solution of the problem is then found by another program, the \emph{solver}. This remedies users from searching in a wide range of possible algorithms for implementing every single QoS or TE problem. This facility makes it possible to efficiently develop  applications using expressive and declarative programming.

The proposed Northbound API is controller independent and does not use specific features of any controller which makes it a more appropriate API for developing QoS routing and TE applications. A set of eight primitives as a new layer of API is also created which constituents an interface for developing QoS routing and TE applications. A few example QoS routing algorithms are implemented in the form of several applications. The performance of these implemented applications are evaluated in various experiments. A series of networks with various sizes and topologies utilised in these experiments. The results of these evaluations indicate a practical performance for even large scale networks. Explanation of the new Northbound API is started with a brief background in Section \ref{Background}, then related works are presented in Section \ref{Related Works}. Section \ref{Architecture} explains the architecture of a routing framework for SDN which includes the proposed Northbound API and Section \ref{Implementation} describes its implementation. Section \ref{Evaluation and Results} discusses the results and evaluation and Section \ref{Conclusion} concludes the paper.\\


\section{Background}\label{Background}
\subsection{Routing Network Flows}
Routing in general, SDN-based or not, involves two entities, network state information and routing algorithms. Network state dynamically changes due to transient load fluctuations, connections in and out and links up and down, so it constitutes a significant part of traffic load and uses resources of the routing devices in many networks. In legacy networks, network state information is gathered  via distributed routing protocols which acquire and distribute it from and to routing devices. In SDN though, the network state information is centrally collected and updated by the controller that directly communicates with routing devices. Consequently, maintaining and updating network state information through SDN controller frees the resources from the network and devices \cite{Quality_of_Service_Routing}.

Routing algorithms are routines or formulas used to make decisions concerning the paths for network flows. The purpose of a routing algorithm is in general determined by the flow's characteristics such as source-destination pairs and the demand as well as services which a provider may want to provide to its customers e.g. guaranteed bandwidth. Since there are finite resources in a network, e.g. capacity, an objective for a network in general is to provide an efficient and fair routing \cite{Network_routing_algorithms_protocols_and_architectures}.

The network flow routing problems are mathematically stated in terms of \emph{graph} notions. Wherever graph is referred in this work, it is assumed \emph{directed}. A directed graph ${\mathcal G(\mathcal N,\mathcal A)}$, is a set of \emph{nodes} $\mathcal N$ and a set of \emph{arcs} $\mathcal A$ which is represented by a set of \emph{ordered} pairs of distinct nodes  $(\mathcal A = \{(u,v)|u \in \mathcal N\ and\ v \in \mathcal N \})$. The number of nodes is $N$ and the number of arcs is $A$. An arc is an ordered pair so, the arc $(u,v)$ is distinguished from the arc $(v,u)$ (arc and link are used equivalently in this text). There are multiple concepts and notions relating to graphs used in this work which are required to be defined before use. 

An arc $(u,v)$ is said outgoing from node $u$ and incoming to the node $v$. The node $u$ is called the \emph{start} node and the node $v$ is called the \emph{end} node of the arc $(u,v)$ and the arc is called \emph{incident} to its \emph{end nodes} $u$ and $v$. The \emph{degree} of a node $u$ is the number of arcs incident to $u$. It is assumed there is only one arc between each pair of nodes in the same direction. An \emph{undirected} graph can be replaced with a directed graph where every undirected arc $(u,v)$ is replaced with two directed arcs $(u,v)$ and $(v,u)$ \cite{Network_optimization_continuous_and_discrete_models}.

\begin{figure}[t]
\centering
\includegraphics [width=5cm, height=5cm, natwidth=500, natheight=550]{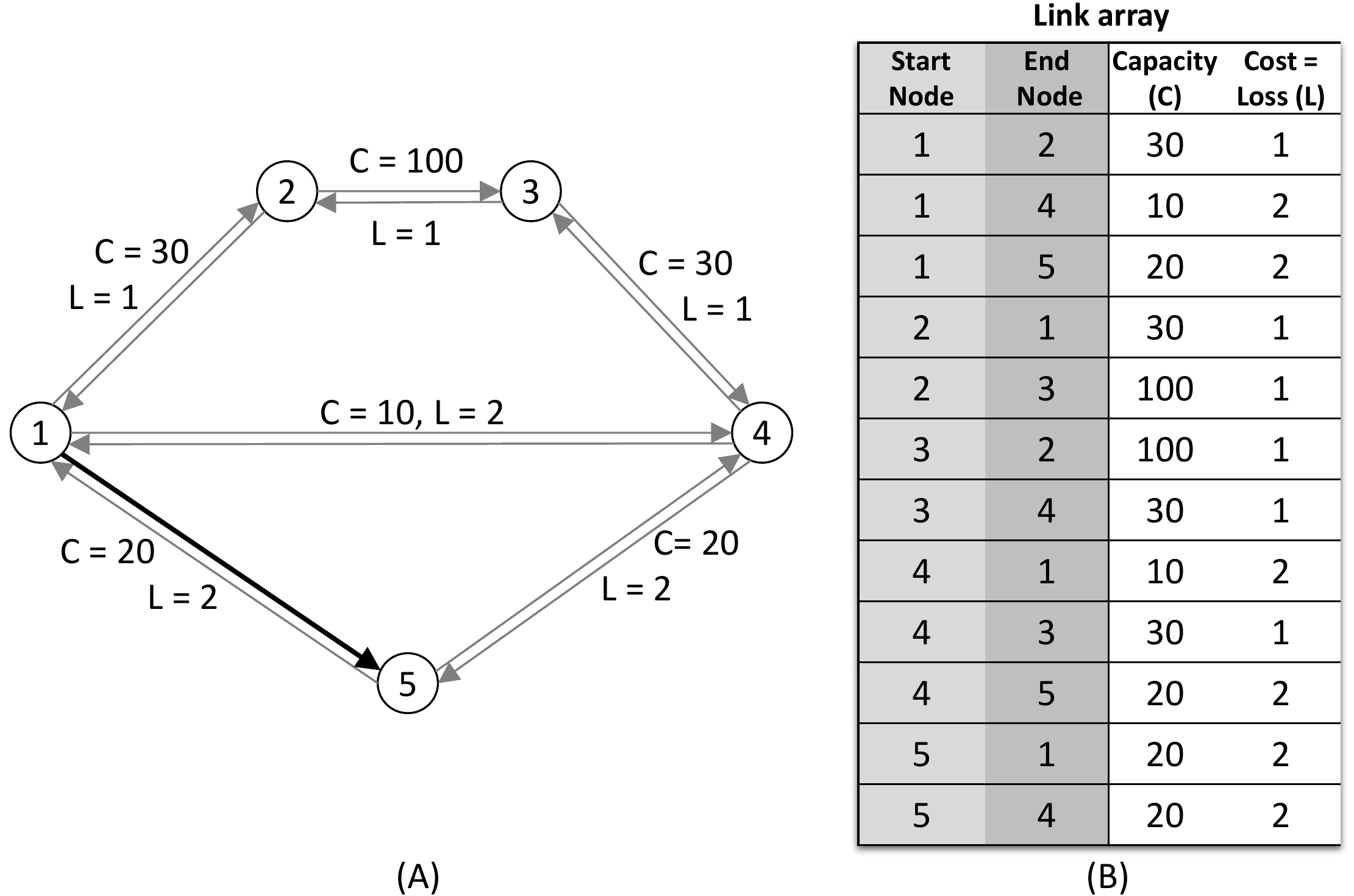}
\caption{A sample network graph and its link array}
\label{fig:link_adjacency}
\end{figure}

For the purpose of programming, a graph (either directed or undirected) can be represented in two ways, an \emph{adjacency matrix} or an \emph{adjacency list}. In the adjacency matrix, graph nodes are numbered from 1 to N and a $N \times N$ matrix $M = (m_{uv})$ represents the graph where

\begin{flalign}
m_{uv} = \begin{cases}
 1\ \ \ \ if\ (u,v) \in \mathcal{A}\\
 0\ \ \ \ \textsl{otherwise}
 \end{cases}
\label{formula:adjacency-matrix}
\end{flalign}

\noindent In the algorithms implemented in this work the adjacency-list representation is used for all the graphs. It is an array $L$ which includes the adjacency lists of all nodes. The adjacency list of a node is the list of all the nodes for which there is an outgoing arc \cite{Introduction_to_algorithms}. This concept is illustrated for a sample network graph in Fig. \ref{fig:link_adjacency}-A with five nodes and six bidirectional links (bidirectional links are represented as two unidirectional links and so there are 12 unidirectional links). The adjacency list of the graph is shown in the first two columns of the table in Fig. \ref{fig:link_adjacency}-B and the two next columns are other link parameters that will be explained and used in the later sections. The adjacency list for the node number $1$, for instance, contains all the nodes for which there is an outgoing arc from the node number $1$ that includes nodes $2$, $4$ and $5$. Node number 1 is considered in the adjacency lists of these nodes, for the purpose of its incoming links, which are expressed in $4$th, $8$th and $11$th rows.

A \emph{path} $P$ in a graph is defined as a sequence of nodes $(n_1,n_2,...,n_k)$ with its corresponding sequence of $(a_1, a_2,...,a_{k-1})$ arcs, where $k \geq 2$. In this arc sequence the $i$th arc, $a_i$, is either $(n_i, n_{i+1})$ which is called \emph{forward arc} or $(n_{i+1}, n_i)$ that is called \emph{backward arc}. Nodes $n_1$ and $n_k$ are called \emph{start node} (or origin) and \emph{end node} (or destination) of the path $P$. A path is called \emph{simple path} if it includes no repeated arcs and nodes except the origin and destination nodes. It is important to know that a sequence of nodes by itself (without its corresponding sequence of arcs) does not specify a path unless the path is either \emph{forward} (all arcs forward) or \emph{backward} (all arcs backward). In all usage of the path in this work, it is assumed forward (unless it is stated otherwise) and a sequence of nodes (or equivalently a sequence of arcs) adequately identifies a path.

A \emph{flow} is a variable to measure the quantity of what is flowing through an \emph{arc}. The flow of an arc $(u,v)$ can be stated mathematically by a real number $f_{uv}$ (or $f(u,v)$)  that is assumed non-negative without loss of generality. There is another concept related to flows in the network which is called \emph{divergence}. The divergence of a node ${u}$ is the total flow exiting the node less the total flow entering the node \cite{Network_optimization_continuous_and_discrete_models}

\begin{flalign}
y_u = \sum_{\{v|(u,v) \in  \mathcal A\}}{f(u,v)} - \sum_{\{v|(u,v) \in  \mathcal A\}}{f(v,u)}\ \ \ \ \ \
\forall u \in \mathcal N\
\label{formula:divergence}
\end{flalign}\\

In order to explain the \emph{flow path} in a network, the \emph{flow vector} concept is used. A flow vector represents a \emph{set of flows} ${f = \{f(u,v)|(u,v) \in \mathcal A\}}$ in the graph $\mathcal G$. A \emph{simple path flow} is a flow vector , i.e. set of arc flows, that is equivalent to send a (positive) flow along that simple path. For the simple path $P$ it is mathematically stated by a flow vector $f$ 

\begin{flalign}
f(u,v) = \begin{cases}
			\ \ a\ \ \ \ if\ (u,v) \in P^+\\
			-a\ \  \ \ if\ (u,v) \in P^-\\
			\ \ 0\ \ \ \ \textsl{otherwise}
 \end{cases}
\label{formula:simple_path_flow}
\end{flalign}\\

\noindent where $a$ is a positive scalar and $P^+$ and $P^-$ are forward and backward subsets of arcs of $P$ respectively. A flow vector $f$ can have two distinguished nodes the \emph{source} and \emph{sink} which have positive and negative divergence respectively. A path $P$ is said \emph{conformed} to a flow vector $f$ if the flow for all forward arcs is positive and it is negative for all backward arcs, and the path's start and end nodes are the source and the sink of the flow vector f respectively, if they are not the same.

The flow vectors studied in this work are constrained by two constraints. The first constraint is called the `\emph{capacity constraint}' which limits the flows to a lower and upper bounds as \cite{Network_optimization_continuous_and_discrete_models}
\begin{flalign}
b(u,v) \leq f(u,v) \leq c(u,v)\ \ \ \ \ \ \forall (u,v) \in \mathcal A\
\label{formula:flow_boundss}
\end{flalign}

\noindent in which $b(u,v)$ and $c(u,v)$ are called the \emph{flow bounds} of the arc $(u,v)$. The second constraint is called the `\emph{conservation of flow}' that states \cite{Network_optimization_continuous_and_discrete_models}
\begin{flalign}
\sum_{\{v|(u,v) \in  \mathcal A\}}{f(u,v)} - \sum_{\{v|(u,v) \in  \mathcal A\}}{f(v,u)}= s_u\ \ \ \ \ \
\forall u \in \mathcal N\
\label{formula:flow_conservation}
\end{flalign}

\noindent In the above equation if $s_u > 0$, it is called the \emph{supply} of the node $u$, and if ${s_u < 0}$, the scalar $-s_u$ is called the \emph{demand} of the node $u$. The flow conservation simply states that the total flows exiting a node is equal to flows entering the node unless the node is either a flow source or sink. A flow vector satisfying capacity and flow conservation constraints is called \emph{feasible flow vector}. 

The capacity of an arc $(u,v)$ is a non-negative real number $c_{uv}$ (or $c(u,v)$) that is the upper flow bound of the arc. The capacity for the graph $\mathcal G$ can be defined as a function ${c:\mathcal A \rightarrow \mathbb{R}_0^+}$ that maps the graph's arcs (ordered pairs) to the non-negative real numbers. If there are two distinguished nodes in a graph $\mathcal G$ as the flow source `$s$' and sink `$t$' then the graph $\mathcal G$ along with its capacity function $c$ is a `\emph{flow network}' $(\mathcal G, c,s,t)$ \cite{Network_flow_algorithms}.

The problem in a routing algorithm usually consists of finding a path from a flow source to a flow sink that optimises a parameter, e.g. minimising a linear cost function, subject to some constraints. Some routing algorithms require to satisfy a condition on a single or multiple parameters rather than optimising them. If the constraint satisfaction or optimisation in finding paths include parameters to satisfy the QoS, the routing algorithm is called the QoS routing algorithm \cite{QoS_routing_in_networks_with_inaccurate_information_theory_and_algorithms}. Examples of these algorithms along with their mathematical statements are presented in section \ref{Implementation}.

The concept of the QoS represents the quantitative and / or qualitative performance criteria in contract between service provider and its customers. These requirements are given as a set of constraints on a link, single or multiple path in the network \cite{P-Quality_of_service_routing_for_supporting_multimedia_applications}. Hence, QoS routing is finding routes in the network based on these requirements which is the first step toward achieving end-to-end QoS guarantees \cite{P-Quality_of_Service_Routing_for_Traffic_with_Performance_Guarantees}. The QoS routing problem, i.e. identifying feasible paths that can satisfy the given QoS constraints, constitutes one of the key issues in all QoS-based networking framework such as integrated services (IntServ), diffrentiated services (DiffServ), and multiprotocol label switching (MPLS) \cite{P-An_overview_of_constraint-based_path_selection_algorithms_for_QoS_routing}. The Intserv framework requires a QoS-aware underlying IP routing table to reserve its resources and guarantee some specific QoS constraints. The Diffserv framework utilises the QoS-based routes to ensure a certain service level agreement. These routes can be requested by, for instance a network administrator, for TE objectives. In MPLS the route selection, possibly subject to QoS constraints, and resource reservation along the path, is done by the source node. The significance of the constraint-based routing algorithms is clearly indicated in these examples as some of the major networking frameworks \cite{Quality_of_Service_Routing}.


\subsection{Constraint Programming}
Constraint programming techniques were initially introduced in the 1960s and 1970s in artificial intelligence and computer graphics \cite{Constraint_programming}. Then they have found applications in many fields such as operations research, programming languages and databases \cite{Handbook_of_constraint_programming}. They have been successfully used to solve many practical problems in a number of fields such as scheduling of air traffic, network security, vehicle routing, planning, chemistry, biology and bioinformatics \cite{Handbook_of_constraint_programming,Constraint_Programming_and_Decision_Making}. Nonetheless, these techniques have potential to be applied in a further range of real-life applications and problems.

The main idea behind CP is to separate stating the problem from its solution. So,  users are only required to state the problem and the solution is found by the general purpose constraint solvers which are designed for this purpose. This allows declarative descriptions to generate very flexible modelling that can be used to effectively solve large, particularly combinatorial problems \cite{Constraint_Programming_In_Pursuit_of_the_Holy_Grail}. 

In order to have a constraint solver to solve a real world problem, it must be stated in the form of a CP \emph{model}. A CP model includes at least three parts
\begin{itemize}
\item \emph{Decision variables}  that represent tasks, metrics or resources of a real world problem.
\item \emph{Variable domains} that are a finite set of possible values for each decision variable.
\item \emph{Constraints} that state the relations (conditions, limitations, properties and bounds) between decision variables.
\end{itemize}
\noindent The constraints in fact restrict the values that all decision variables can have at the same time \cite{Constraint_Programming_In_Pursuit_of_the_Holy_Grail}. 

As an example, consider the \emph{Shortest Path} routing algorithm. Here it is assumed that arcs of the graph are associated with a scalar number $a(u,v)$ which represents the \emph{cost} and the cost of a forward path is sum of the costs of its arcs. Then for a given pair of source-sink nodes the problem is to find a forward path with minimum cost that connects the source to sink. It can be stated mathematically as \cite{Network_optimization_continuous_and_discrete_models}

\[\textsl{minimize}\ \ \ \ \sum_{(u,v) \in \mathcal{A}}{a(u,v)f(u,v)}\]
subject to
\begin{flalign}
\label{formula:shortest_path_1}
\sum_{\{v|(u,v) \in  \mathcal A\}}{f(u,v)} - \sum_{\{v|(u,v) \in  \mathcal A\}}{f(v,u)}= 
	\begin{cases}
		\ \ 1\ \ if\ u=s,\\
		-1 \ \ if\ u=t,\\
		\ \ 0\ \textsl{otherwise}
	\end{cases}
\end{flalign}\\

\noindent Then the vector $f$ of the form 

\begin{flalign}
f(u,v) = \begin{cases}
			 1\ \ \ \ if\ (u,v) \in P\\
			 0\ \ \ \ \textsl{otherwise}
 \end{cases}			
\forall (u,v) \in \mathcal{A}
\end{flalign}\\

\noindent is a feasible flow vector for the problem (\ref{formula:shortest_path_1}) and the cost of $f$ is equal to the length of $P$. The  problem can be stated as a CP model as follows:

\begin{itemize}
\item \emph{\textbf{Decision Variables}}:\\
$P\{n_i\} :$ represents flow membership in which\\
$\{n_i|\ i\ \textsl{in}\ [1\ N]\}$ represents the network nodes and\\
\textsl{Domain}${[P] =\{0,1\}}$.
\item \emph{\textbf{Constraints}}:\\
$\textsl{for}\ i\ \textsl{in}\ [1\ N]:\\$
\[
    \left\{
           \begin{array}{ll}
           
P\ \textsl{includes no repeated}\ n_i\ \\ 
\textsl{Sum}[Flow\_in[n_i]] -  \textsl{Sum}[Flow\_out[n_i]]=\alpha\\ 
\alpha=\ \textsl{1}\,\ \ \ \ \textsl{if}\ \ \ i\ \ \textsl{is}\ \ \textsl{Source}\\ 
\alpha=\textsl{-1}\,\ \ \ \ \textsl{if}\ \ \ i\ \ \textsl{is}\ \ \textsl{Sink}\\ 
\alpha=\ \textsl{0}\,\ \ \ \ \textsl{else} 

\end{array} \right.
\]
  
\item \emph{\textbf{Objective}}: \\
\textsl{minimize Length}$[P]$\\
\end{itemize}

The solution of a CP model is the allocation of values for the decision variables from their domains that simultaneously satisfy all the constraints. It might be possible to find 
\begin{itemize}
\item a single solution with no priority
\item all solutions 
\item an optimal solution, given an \emph{objective function}
\end{itemize}

\noindent If there exist such an objective function in terms of some or all decision variables that identifies an optimal solution (similar to the above model), the problem is called an \emph{Optimisation Problem (OP)} rather than being called a \emph{Constraint Satisfaction Problem (CSP)} \cite{Constraint_Programming_In_Pursuit_of_the_Holy_Grail}. The solvers enumerate possible variable-value combinations intelligently and search the solution space either systematically or through some forms of complete or incomplete searches. The performance of these search methods depends on the statement of the problem \cite{Handbook_of_constraint_programming}.\\

\begin{figure}[!b]
\centering
\includegraphics [width=7.5cm, height=6cm, natwidth=1000, natheight=500]{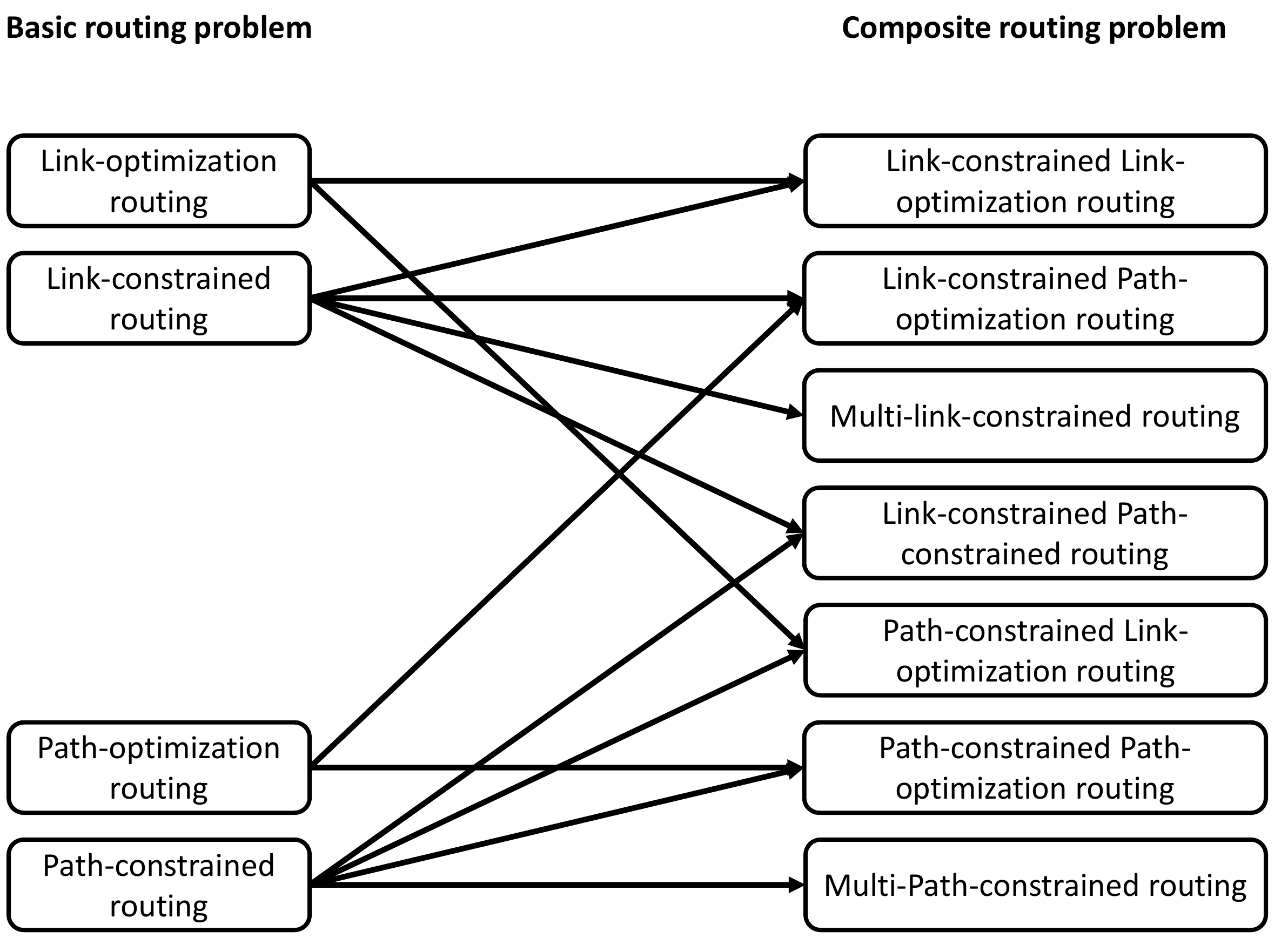}
\caption{Basic and composite QoS routing problems \cite{P-An_overview_of_quality_of_service_routing_for_next_generation_high_speed_networks_problems_and_solutions}}
\label{fig:QoS-routings-classes}
\end{figure}

\section{Related Works}\label{Related Works}
There are many works related to QoS routing problems in the legacy networks. One of the best classifications of the QoS routing problems is presented in \cite{P-An_overview_of_quality_of_service_routing_for_next_generation_high_speed_networks_problems_and_solutions} which is shown in Fig. \ref{fig:QoS-routings-classes}. It divides the unicast routing problems into two major categories, \emph{basic} and \emph{composite} routing problems. The criteria used for this classification is the \emph{number of QoS metrics} applied in the problem. A basic routing problem includes only a single QoS metric such as delay, jitter or bandwidth. The routing problem in this case might include the \emph{metric optimisation}, such as minimum delay path, or it might include \emph{metric constraint}, such as bandwidth constrained path. In addition, the metric might be related to a \emph{link parameter}, e.g. bandwidth, or it might be related to a \emph{path parameter} such as delay. Hence, there are four types of the basic unicast routing problems as seen in Fig. \ref{fig:QoS-routings-classes}. The composite routing problems can be used to define multi-constraint QoS routing problems. They can be expressed as the combination of a single-metric-optimisation problem with one or more metric-constraint problems. Some multi-constraint QoS routing problems can be expressed as the combination of two or more metric-constraint problems. Depending on the metric type, is a link or path metric, several types of composite routing problems are created which correspond to various proposed QoS routing problems.

In the works related to QoS routing in SDN, the Northbound API is part of SDN architecture that is expected to provide related facilities. The Northbound and Southbound APIs are the two key elements of SDN framework which provide communication between their below and above layers. While there is a widely accepted proposal for the Southbound API, i.e. OpenFlow, no common proposal is accepted for the Northbound API. Despite the variety of the available Northbound APIs, little work is done on addressing the QoS routing in these platforms. In fact, there is no Northbound API to implement abstractions required for implementing all the basic and composite QoS routing algorithms, similar to legacy networks.

Some existing controllers such as Floodlight \cite{Floodlight}, NOX \cite{NOX_towards_an_operating_system_for_networks} and OpenDaylight \cite{OpenDaylight} have proposed and defined their own Northbound APIs with the specific definitions. They tend to deal with the packet and port level manipulations rather than routing and QoS routing abstractions. Even the QoS module provided by the Floodlight controller mostly deals with the hardware, packet level details and low level configurations. This makes the proposed Northbound interfaces unsuitable choice for developing QoS routing applications that require higher levels of abstractions to implement QoS routing concepts (such as capacity constraint in the path selection).

Some programming languages such as Frenetic \cite{Frenetic_a_network_programming_language}, Pyretic \cite{Modular_sdn_programming_with_pyretic} and Hierarchical Flow Tables (HFTs) \cite{Hierarchical_policies_for_software_defined_networks} also provide an abstraction of the controller functions and data plane behavior for the upper layer. Pyretic, Frentic and HFTs are examples of declarative Northbound programming languages with facilities for expressing packet forwarding policies. They also provide facilities for parallel and sequential composition of various non-overlapping  modules. Despite the abstractions and facilities created for applying packet forwarding policies and module composition, they still work at the packet level and no high level abstraction of QoS functions is provided yet. Furthermore, implementing routing and QoS routing algorithms in these languages is done through the same algorithms that is currently applied in legacy networks, e.g Dijkstra algorithm for the shortest path routing in Pyretic \cite{Extending_Dijkstra_shortest_path_algorithm_for_SDN}. This makes developing QoS routing and TE applications a cumbersome job, that requires users be familiar with programming techniques for efficient implementation of algorithms in possibly hundreds of lines of code.

The PANE controller  \cite{Participatory_networking_an_API_for_application_control_of_SDNs} provides an API that allows applications such as VoIP and video to request network resources and reserve bandwidth for specific QoS requirements. It truly recognises the need for an API between the network’s control-plane and its users, applications, and end-hosts.  Though, the functionalities that PANE provide is limited to a few features provided in the PANE controller and users are not able to develop QoS routing and TE applications. As such, it can be considered an application rather than a platform for developing networking applications.

SFNet \cite{Towards_software-friendly_networks} is another Northbound API that emphasises on the closer interaction between the applications and network. It provides some network services to the applications through a high-level API that hides the details of the implementation. It suggests a software friendly networking (SFNet) paradigm in which it is possible for the applications to reserve bandwidth or query network congestion state. The main idea of SFNet, i.e. to create abstractions of the network services, is one of the core ideas of SDN, the implementation of SFNet can not be considered optimal. It has also a very limited scope that focuses on services such as bandwidth reservation and queries to request the network's congestion state. The mere functionality of the QoS component of SFNet is to reserves the bandwidth along the shortest path between two network nodes. This can clearly be insufficient or useless in many situations in which the shortest path lacks the enough bandwidth.

While the main purpose of SDN is to enhance innovation in the network, a platform for the efficient implementation of  present and new networking ideas is a clear gap in this field. This is specially the case with QoS routing and TE applications. The above Northbound APIs and programming languages, despite providing some features and facilities, cannot be considered the desired platform. Most of the available Northbound APIs are dealing with the lower level details such as packet and hardware manipulation. Only a few Northbound APIs such as the QoS module of the Floodlight controller provide abstractions for QoS functionalities that is still dealing with the packet and hardware level. Some of these works such as PANE are more considered application, rather than a Northbound API or a platform for developing applications. Some of them like SFNet have a limited scope which is far from a platform that can be used to address real QoS routing requirements of today's networks.

Due to these issues, implementing QoS routing applications in these Northbound APIs and programming languages is inconvenient, similar to developing desktop applications in Assembly language. The users not only have to find an efficient algorithm to solve each individual QoS routing problem, but they also must implement it through programming languages such as Python, JAVA and C which might require hundreds of lines of code. As such, it necessitates both theoretical knowledge of the competent algorithms for each QoS routing application and the ability for its efficient implementation. Furthermore, since these Northbound APIs are mostly compatible with their corresponding controller, applications are needed to reimplemented for each new controller.

DEFO is a very recently presented approach to control forwarding paths in \emph{non-SDN} carrier grade networks. It uses a declarative and expressive language to efficiently address many QoS routing requirements. It has a two-layer architecture that separates connectivity and optimization tasks. The high-level goals expressed in the declarative language is translated into compliant network configurations by the centralized optimizer. The results of the evaluation of DEFO indicates it achieves better trade-offs for the classic QoS goals and supports a larger set of goals. DEFO has been able to optimise large ISP networks in a few seconds for the classic QoS goals which required more than one night on a powerful server through linear programming. DEFO relies on CP and new heuristics in order to achieve these advantages in the scalability and flexibility \cite{A_Declarative_and_Expressive_Approach_to_Control_Forwarding_Paths_in_Carrier-Grade_Networks}. Although DEFO is not a Northbound API in the SDN architecture, there are lessons to be learned from its approach. In fact, it includes many features of a desired Northbound API for efficient addressing of QoS routing and TE requirements in the network.

The main advantages of DEFO are achieved by using CP techniques for the network optimisation. This is not the first time CP is suggested for the network optimisation objectives. Previously, it has been suggested in  \cite{Challenges_for_Constraint_Programming_in_Networking} for efficient solving of various problems in legacy networks such as TE and network design.  The same idea can be applied in SDN which provides more facilities and freedom for the implementation of CP in the network optimisation. Since the task of a QoS routing algorithm is to find paths subject to a single or multiple constraints, CP  is an evident choice for solving these problems.  

Using CP in a Northbound API makes it possible to develop QoS routing applications without the requirement of knowledge of various algorithms to solve each QoS routing problem. Users state the problem in terms of CP variables and constraints and the solver finds the solution without further intervention from the user. While the simplicity of stating the problems make it a really useful tool for developing applications, the presence of solvers with strong intelligent search techniques ensures the efficiency is not less than other available solutions. Indeed, the DEFO example indicates it is even much faster than similar methods for some QoS routing and TE applications.

A CP-based Northbound API, hence, can address the current gap of an efficient application development platform in SDN.   
The QoS routing functionalities / constraints, due to the nature of the problem, can easily be stated as some CP primitives. These primitives can model basic QoS routing problems and their combinations can be used to express composite QoS routing problems. This is the methodology used in this work to design and implement the new Northbound API, SCOR. A routing framework is also proposed in this work which includes other elements required by the QoS routing applications. The design and architecture of the proposed solution is discussed in the next section.\\

\section{Design}\label{Design}
\subsection{Architecture}\label{Architecture}
\begin{figure*}[!t]
\centering
\includegraphics [width=18cm, height=9cm, natwidth=1000, natheight=500]{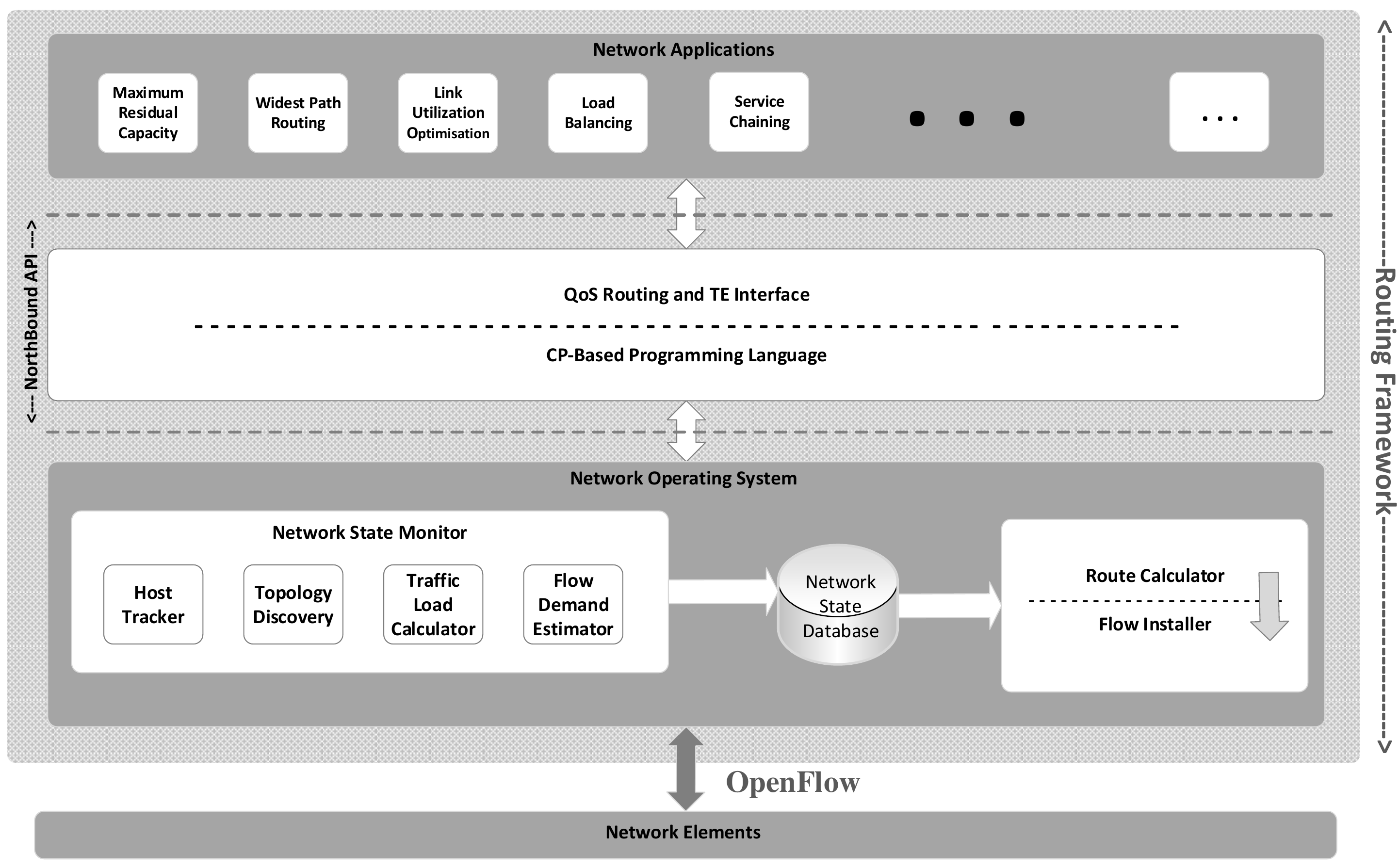}
\caption{The routing framework based on SCOR}
\label{fig:Framework}
\end{figure*}

Fig. \ref{fig:Framework} shows the structure of a routing framework based on the proposed Northbound API. It has three layers as the \emph{Network Operating System (NOS)}, the \emph{Northbound API} and the \emph{Network Applications}. The bottom layer of the framework, NOS, directly communicates with the \emph{Network Elements} through common protocols such as OpenFlow. It uses the three components \emph{Network State Monitor}, \emph{Network State Database} and \emph{Route Calculator} to acquire network state such as traffic loads and flow demands from Network Elements and provide it for the applications.

The Network State Monitor is the component of NOS that collects the network state information by communicating with the Network Elements (layer). It includes four modules as \emph{Host Tracker}, \emph{Topology Discovery}, \emph{Traffic Load Calculator} and \emph{Flow Demand Estimator}. The Host Tracker and Topology Discovery are two standard modules present in many controllers which provide network topology information. The Traffic Load Calculator module periodically reads the flow statistic of ports and keeps track of network links' utilisation and available network resources \cite{Efficient_topology_discovery_in_OpenFlow-based_Software_Defined_Networks}. The last module, Flow Demand Estimator, provides periodic estimations of the flow demands in the network inputs. 

The network state information provided by the above component is recorded in the Network State Database component which is queried by the Network Applications through the Route Calculator component. This information is needed by any network application that modifies routing or other parameters of the network. The Route calculator also receives network routes from the applications and translate them to flow rules. These flow rules are then installed in the network elements through a sub-module, the \emph{Flow Installer}.

The next layer up is the proposed constraint programming-based Northbound API, SCOR. It is explained in more details in the next sub-section as the main contribution of this work. The top layer of the framework is the Network Applications. This layer includes different applications that are created in the SCOR. The range might include common QoS routing and TE applications such as maximum bandwidth path, maximum residual capacity path, link utilization optimization, and load balancing. It might also include applications that implement newer services such as service chaining. These applications get the network topology and state information from the NOS through route calculator component, calculate the route, and return it to route calculator.\\

\subsection{QoS Routing and TE Interface}\label{QoS Routing and TE Interface}
Purposing to use CP in SDN and designing the required primitives is the main contribution of this work. This includes designing the Northbound API and an interface for developing QoS routing and TE applications. As it seen in Fig. \ref{fig:Framework}, SCOR consists of two sub-layers which are equivalent to two levels of programming. The bottom level is the \emph{CP-based Programming Language} and the higher level is the set of primitives which forms the \emph{QoS Routing and TE Interface}. The bottom sub-layer, CP-based programming language, can be considered as a general platform for developing routing and other types of applications. CP techniques have been shown efficiently fit to address many problems in network design, planning and analysis \cite{Challenges_for_Constraint_Programming_in_Networking}. This CP-based programming language enables users to state their problems in a wide range of networking domains by focusing on `\emph{what should be done}' rather than `\emph{how they should be done}'.

The second sub-layer, QoS routing and TE interface, is particularly designed to address QoS routing and TE requirements. It is consisted of two groups of predicates which totally include eight \emph{predicates} as listed in Table \ref{tab:8-predicates}. The first group includes predicates 1 to 5 which are designed to implement the single and multi-metric QoS routing problems. They implement various constraints such as the flow conservation, capacity constraint and cost, and they are fitted to realise both basic and composite routing problems. 
 
The first and the most fundamental routing concept to realise is the flow path. Theoretically, the flow path is defined as a set of arc flows that complies with two constraints, flow conservation rule (Eq. \ref{formula:flow_conservation}) and the capacity constraint (Eq. \ref{formula:flow_boundss}). Though, in general it is possible to define a flow path that does not satisfy the capacity constraint. As such, the \emph{Network Path} predicate applies only the first rule, the flow conservation, to identify a flow path and the capacity constraint is designed as a separate predicate, the \emph{Link Capacity Constraint}. Combination of these two predicates can identify a feasible flow path (Section \ref{Background}) in a flow network.

\begin{table}[!t]
\renewcommand{\arraystretch}{1.6}
\caption{Set of predicates forming QoS routing interface}
\vspace*{-2mm}
\label{tab:8-predicates}
\begin{tabular}{|c|p{2cm}|p{5cm}|}
\hline
\textbf{Item}  & \textbf{ Predicate Name}  & \textbf{Functionality}     \\ \hline
$1$ & \centering \emph{Network Path}  &  Finds all the paths from a flow source `s' to a flow sink `t' in the given network graph\\ \hline
$2$ & \centering \emph{Link Capacity Constraint}  & Places the flows on arcs that can accommodate their demands \\ \hline
$3$ & \centering \emph{Residual Capacity} & Calculates the residual capacity of arcs, after placing all the flow demands \\ \hline
$4$ & \centering \emph{Path Capacity Constraint} &  Removes all arcs with a capacity less than a specified value, from the network graph \\ \hline
$5$ & \centering \emph{Path Cost} & Calculates the total cost of a path based on the cost of arcs forming the path  \\ \hline
$6$ & \centering \emph{Delay} & Calculates the queueing delay for the given network graph and flows  \\ \hline
$7$ & \centering \emph{Congestion} & Calculates the congestion for the given network graph and flows  \\ \hline
$8$ & \centering \emph{Link Utilisation} & Calculates the link utilisation for the given network graph and flows  \\ \hline
 
\end{tabular}
\end{table}


The next two predicates, the \emph{Residual Capacity} and \emph{Path Capacity Constraint} create abstractions relating to capacity that are frequently used in the QoS routing applications. The former finds the value of the residual capacity of all arcs in the network graph, assuming all the known flow demands are routed. The latter removes arcs with a capacity less than a specified limit from the network graph. This is particularly useful when calculating routes with two or more metrics including the bandwidth or capacity limitation. while this predicate applies the capacity constraint, it reduces the network graph and makes imposing the next constraints faster.

The path cost is the other concept required for the implementation of many QoS routing problems. The \emph{Path Cost} predicate which realises this constraint assumes the cost of a path is the sum of its arcs (links) costs. The cost metric can be different network parameters such as \emph{number of hops}, links delay (transmission and propagation) or the link lease price. Current routing protocols use various network parameters as distance or cost metric. In Open Shortest Path First (OSPF) protocol it is possible to use either of throughput of the links, link availability or Round-Trip Time (RTT) as the cost. The Routing Information Protocol (RIP) uses the number of hops to the destination and Interior Gateway Routing Protocol (IGRP) uses a combination of node delay and available bandwidth. 

There are two cases of routing algorithms that use multiple cost metrics at the same time. In the first case, one cost metric is optimised and the rest are constrained to satisfy some limits, which is an optimisation problem. In the second case, all cost metrics are constrained to satisfy their corresponding limits, that is a constraint satisfaction problem. As an example consider a routing algorithm that finds the path with minimum RTT and the link availability greater than specific value. Both metrics can be used as a cost metric and the problem is a multi-cost-metric routing problem. SCOR can easily implement these algorithms by assigning the cost metrics to different decision variables and applying the Path Cost predicate for each one separately.  

Routing problems that are addressed by the group of five above predicates include all  problems in which one or more static network parameters are either optimised or constrained as stated in Section \ref{Related Works}. Though, there are still other TE problems in which the desired network parameters dynamically changes with the traffic. For instance, the queueing delay in each network node, not only depends on the capacity of network paths, but it also depends on the amount of traffic flowing into the network. It can be compared to the transmission delay of links which is a static parameter depending only on the capacity of the links. In this way, another group of predicates are needed to realise these concepts.

Accordingly, the second group of predicates which includes \emph{Delay}, \emph{Congestion} and \emph{Link utilisation} realises constraints required to implement advanced TE applications such as link utilisation optimisation, maximum concurrent flow and network load balancing. All these predicates are based on the concept of the residual capacity which is implemented as a separate predicate. They use the Residual Capacity predicate and illustrate the \emph{Nested Predicate} feature as a technique in efficient code re-usability. Though, it is not limited to these predicates and it is also applied in the predicates of the first group. These are explained thoroughly in next section which demonstrates the implementation of the Northbound API and also some applications created through it. \\
 
\section{Implementation}\label{Implementation}
\subsection{Routing Framework}
The routing framework is currently implemented based on the POX controller. Nonetheless, it does not bound to any specific feature in POX and can be easily adopted for the other controllers. This is specially the case with the ONOS \cite{ONOS_towards_an_open_distributed_SDN_OS} and its intent subsystem. Intents are policy-based directives that makes it possible for the applications to state their network requirements in terms of network policies rather than the mechanisms. An intent compiler translates intents, received from the applications, to actionable operations on the network environment. This is very similar to the procedure happening in SCOR. In a similar manner, users / applications state their requirements in SCOR in terms of high level / global network policies without knowing how to do it, and SCOR compile them to installable network flow rules. This makes ONOS and intent framework a strong candidate for the future implementation of this routing framework.

SCOR is implemented in MiniZinc  \cite{MiniZinc_Towards_a_Standard_CP_Modelling_Language}. MiniZinc is a simple declarative CP modelling language. In addition to its pre-packaged solvers, e.g. Gecode \cite{Modeling_and_programming_with_Gecode}, MiniZinc can easily use other available solvers such as Jacop \cite{Jacop-java_constraint_programming_solver} and ECLiPSe \cite{Constraint_logic_programming_usingECLiPSe}. The ease of implementation,  simplicity, expressiveness and compatibility with many solvers has made it a practical choice for the standard CP modelling language. It includes a wealthy library of the \emph{global constraints} which model high-level abstractions. \cite{MiniZinc_Towards_a_Standard_CP_Modelling_Language}. 

A problem is usually stated in MiniZinc as two parts / files, the \emph{model} and \emph{model data}. The model uses variables (including `\emph{parameters}' and `\emph{decision variables}') and constraints to describe the structure of a CP problem. The value of parameters are fixed and can be given when declared or in a separate statement or in a separate file, the model data. The value of decision variables is unfixed and can only be fixed after solving the problem. Indeed, the whole aim of the constraint solving is to fix the value of decision variables. Different model data files can be used with a single model to assign different values for the parameters \cite{Specification_of_MiniZinc}. For instance, the \emph{Least Cost Path} model described in Algorithm \ref{alg:SP-MiniZinc} models the least cost path algorithm for an individual network topology which can be defined through a model data file. Different model data files can be used to apply a single least cost path model on various network topologies.

\begin{figure}[!t]
\centering
\includegraphics [width=7.5cm, height=6cm, natwidth=800, natheight=540]{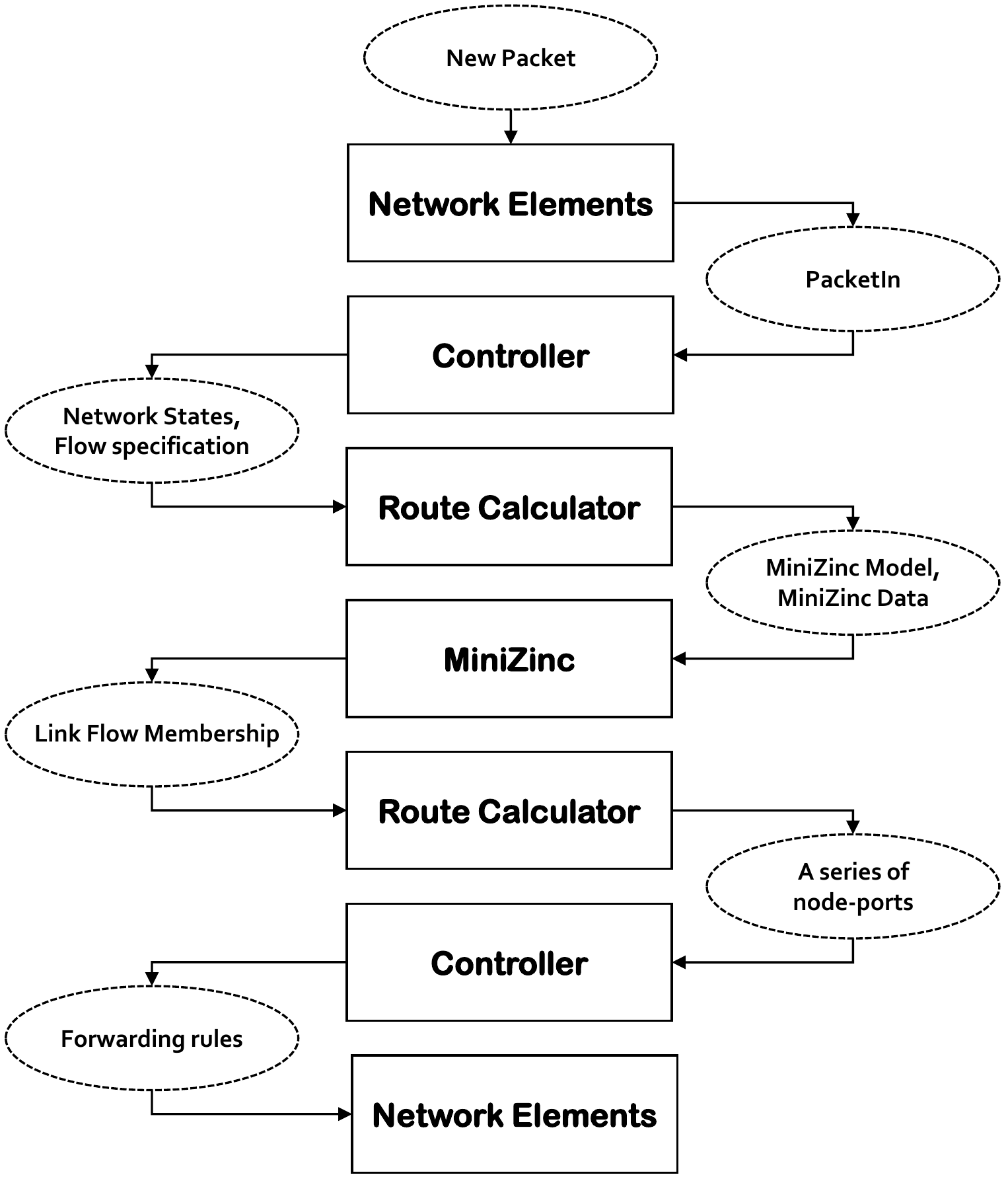}
\caption{The flow of events in the implemented routing framework}
\label{fig:flow_diagram}
\end{figure}

Figure \ref{fig:flow_diagram} shows the flow of events in the implemented routing framework. When a new packet arrives in a network element, if there is no rule installed to specify the required action for that packet, a PacketIn message is sent to the controller. The routing framework, as a module of POX, is listening to the POX core events and receives the PacketIn events. It extracts the flow specifications (network protocol, source and destination addresses, and upper layer source and destination ports) from the packet and passes them along with network state information to the Route Calculator module. The network state collection starts from the beginning of the controller's operation through Topology Discovery and Host Tracker modules of POX that are called in the routing framework. The Route Calculator module converts the received information, the flow and network state, to a MiniZinc data file.

However, MiniZinc is not able to use this data file unless the model file is defined. These models express the users / applications objectives (similar to intents in ONOS) in the form of high level network policies or constraints. In the current implementation, a selected MiniZinc model is defined in the Route Calculator module which is used for each of the experiments explained in Section \ref{Evaluation and Results}. In the next versions it will be possible to apply a model from the set of models automatically selected based on the traffic type or other network requirements. The Route Calculator module runs MiniZinc by calling a chosen CP solver through command line and passing the model and data files as arguments of the command. 

Then the MiniZinc solver finds the solution of the received model-data problem and returns the output in the form of an array of `$0s$' and `$1s$' which is called `\emph{Link Flow Membership}'. A `$1$' value in the Link-Flow-Membership array indicates its corresponding link is determined as part of the \emph{flow path} by the solver, and a `$0$' means it is not on the flow path. The Route Calculator module reads the command line output of MiniZinc, i.e. the Link-Flow-Membership array, and uses the network information to convert this array to a sequence of node-ports which constitutes the flow path. This sequence of node-ports is then passed to the controller which converts it to a series of forwarding rules to be installed on their corresponding network elements. Although the current version of the framework is implemented for calculating the constrained network routes, it is not limited to this application and route calculator can be seen as a general network rule calculator / generator.\\

\begin{table}[!b]
\renewcommand{\arraystretch}{1.6}
\caption{Parameter and variable names used in the predicates}
\label{tab:Predicate_Names}
\centering
\begin{tabular}{|c|p{4cm}|c|}
\hline
\textbf{Name}  & \textbf{Description} & \textbf{Type}    \\ \hline
${N_{nodes}}$  & Number of nodes & Integer \\ \hline
${N_{links}}$ & Number of links & Integer \\ \hline
${N_{flows}}$ & Number of flows & Integer \\ \hline
${Nodes}$ & List of nodes & 1D Array of integers\\ \hline
${Links}$ & List of start and end nodes of links along with their cost and capacity & 2D Array of integers\\ \hline
${LFM}$ & Link-Flow-Membership array & 2D Array of [0,1] \\ \hline
${Flows}$ & List of flow demands & 1D Array of integers\\ \hline
${s}$ & List of flow sources & 1D Array of integers\\ \hline
${d}$ & List of flow sinks (destinations) & 1D Array of integers\\ \hline
${Limits}$ & List of lower bound of capacities & 1D Array of integers\\ \hline

\end{tabular}
\end{table}

\subsection{MiniZinc predicates}
The global constraints of MiniZinc, which state predefined constraints, can be added to models by using an \emph{include} expression (e.g.  \textbf{include} \emph{"globals.mzn";}). New constraints can also be defined and included in the models through \emph{MiniZinc predicates} in a similar manner. MiniZinc predicates that can include one or more constraints, are similar to functions or methods in procedural programming languages. They add more abstraction and modularisation to MiniZinc and make programming the constraints more beautiful.

\begin{table}[!t]
\renewcommand{\arraystretch}{1.6}
\caption{Some common MiniZinc expressions}
\label{tab:MiniZinc-statements}
\centering
\begin{tabular}{|l|c|}
\hline
\textbf{Expression}  & \textbf{Description}     \\ \hline
${constraint}$  & expresses constraints to be satisfied \\ \hline
${include}$ & accessing library \\ \hline
${index\_set}$ & gives the index of a 1D array \\ \hline
${index\_set1of2}$ & gives the index of a 2D array for the 1st dimension \\ \hline
${index\_set2of2}$ & gives the index of a 2D array for the 2nd dimension \\ \hline
${output}$ & expresses what to be printed out \\ \hline
${solve}$ & defines constraint satisfaction or what to be optimized  \\ \hline

\end{tabular}
\end{table}

The predicates explained in this section use a few parameters and decision variables that are listed in Table \ref{tab:Predicate_Names} along with a short description and their types. In order to facilitate user to understand the predicates' code, some common MiniZinc expressions with their descriptions are stated in Table \ref{tab:MiniZinc-statements}. Though, a comprehensive tutorial is available through \cite{A_MiniZinc_Tutorial} that provides syntaxes, descriptions, ample examples and a lot of other helpful information. Since some predicates use other predicates, the order in which predicates are explained is different from Table \ref{tab:8-predicates}.\\

\subsubsection{Network Path Predicate}
All routing applications in SCOR require to identify network paths and then apply other constraints which states their objectives. They rely on this predicate for the network path definition. This predicate is based on Eq. \ref{formula:flow_conservation} that defines a network path as a set of arc flows which comply with the flow conservation rule constraint. In all predicates explained in this section the flows are assumed unsplittable. The MiniZinc code for this predicate is shown in Algorithm \ref{alg:Path-Predicate-MiniZinc}. Parameter and variable names are as explained in Table \ref{tab:Predicate_Names}. The $Links$ array not only includes the adjacency list of the network (Section \ref{Background}), but it also includes other parameters of the links such as the capacity and cost (see Fig. \ref{fig:link_adjacency} for an example). 

\begin{algorithm}[!b]
\caption{The Network Path predicate, to calculate network paths }
\label{alg:Path-Predicate-MiniZinc}

\begin{small}
${1: \textbf{forall}(i\ \textbf{in}\ 1..N_{nodes})(}$\newline
${2: \ \ \textbf{forall}(j\ \textbf{in}\ 1..N_{flows})(}$\newline
${3: \ \ \ \ \textbf{sum}(k\ \textbf{in}\ 1..N_{links}\ }$ 
${\textbf{where}\ Links[k,2]=i)(LFM[k,j]) + }$ \newline
${4: \ \ \ \ (\textbf{if}\ i\ =\ d[j]\ \textbf{then}\
1\ \textbf{else}\ 0\ \textbf{endif})\ =}$ \newline
${5: \ \ \ \ \textbf{sum}(k\ \textbf{in}\ 1..N_{links}\ }$
${\textbf{where} \ Links[k,1]=i)(LFM[k,j])+ }$\newline
${6: \ \ \ \ (\textbf{if}\ i\ =\ s[j]\ \textbf{then}\ 
1\ \textbf{else}\ 0\ \textbf{endif})}$ \newline
${\ \ \ \ \ \ \ \ \ \ \ \ \ \ \ \ \ \ \ \ \ \ \ \ \ \ \ \ \ \ \ \ \ )}$ \newline
${\ \ \ \ \ \ \ \ \ \ \ \ \ \ \ \ \ \ \ \ \ \ \ \ \ \ \ \ \ \ \ )}$
\end{small}

\end{algorithm}


The decision variable in this predicate is the $LFM$, Link-Flow-Membership array. The number of rows of LFM is equal to the numbers of rows of the $Links$ array (i.e. $N_{links}$) and the number of its columns is equal to the number of concurrent flows (i.e. $N_{flows}$). For instance, $LFM[10,1] = 0$ indicates that link number $10$ (i.e. the $10$th row of the $Links$ array, e.g. the link from node number $4$ to node $5$ in  Fig. \ref{fig:link_adjacency}) does not belong to the path of flow number $1$. The objective is to finalise the value of the decision variable. It means to determine links and flows relationship i.e which links belong to the set of links that form a path for each flow. 

The Network Path predicate enforces the unity flow conservation rule (Section \ref{Background}) for each flow in which a unit flow is generated in the flow source node, `$s[j]$', and sinked in a flow destination node, `$d[j]$'. This necessitates sum of the flows entering each node $i$ (line 3, $Links[k,2]=i$ means wherever the node $i$ is the end node of the arc) to be equal to flows exiting the node $i$ (line 5, $Links[k,1]=i$ means wherever the node $i$ is the start node of the arc), unless the node $i$ is either the source of $j$th flow (line 6, $i=s[j]$), or its destination (line 4, $i=d[j]$). The predicate investigates this rule at each node and generates an equation that states the above summations of flows entering in and exiting from the node. The decision variable, $LFM$, is finalised after solving the resulting system of equations from all nodes in the solver. \\

\subsubsection{Path Capacity Constraint Predicate}
The main purpose of this predicate is to reduce the network graph which might speed up the computation or enforcement of another constraint. Whenever there is a lower limit on the link capacities, this predicate can be used as an initial stage before applying a next time consuming procedures. This predicate removes all the links with capacities less than a specified value from the network graph. Hence, the resulting graph is smaller than initial network and applying a new constraint is faster. The MiniZinc code for this predicate is shown in Algorithm \ref{alg:Path-constraint-Predicate} where the decision variable is the $LFM$. It investigates the capacity of all links ($Links[i,3]$) regarding the lower bounds on capacities associated with all flows ($Limits[j]$). Using separate values for each flow is preferred as it make it possible to apply it on multiple concurrent flows. When a link $i$ has a capacity less than the associated limit of a flow $j$, it removes the link from the graph of that flow by assigning zero for the corresponding Link-Flow-Membership entry (i.e. ${LFM[i,j] = 0}$ line 3). \\

\begin{algorithm}[!b]
\caption{The Path Capacity Constraint predicate, to remove links with capacities less than specified limits}
\label{alg:Path-constraint-Predicate}

\begin{small}
${1: \textbf{forall}(i\ \textbf{in}\ 1..N_{links})(}$\newline
${2: \ \ \textbf{forall}(j\ \textbf{in}\ 1..N_{flows})(}$\newline
${3: \ \ \ \ \textbf{if}\ Links[i,3] < Limits[j]\ \textbf{then}\ LFM[i,j] = 0}$\newline 
${4: \ \ \ \ \textbf{else}\ true\ \textbf{endif}}$ \newline
${\ \ \ \ \ \ \ \ \ \ \ \ \ \ \ \ \ \ \ \ \ \ \ \ \ \ \ \ \ \ \ \ \ )}$\newline
${\ \ \ \ \ \ \ \ \ \ \ \ \ \ \ \ \ \ \ \ \ \ \ \ \ \ \ \ \ \ )}$
\end{small}

\end{algorithm}

\subsubsection{Residual Capacity Predicate}
This predicate computes the \emph{residual network}, i.e the network in which capacity of the links is reduced by the value of flows passing through them. In other words, it computes the residual capacity of all links when all flows are placed (paths for all flows are determined). The residual capacity is a common concept in many congestion-related routing algorithms. When there are multiple concurrent flows, the order of placing flows changes the residual network and the condition for placing the next flows. In this case where capacity constraints (Eq. \ref{formula:flow_boundss}) should be satisfied by a flow, each next flow can only be placed in a link if there is enough residual capacity on the link. This creates a dynamic variable which not only depends on the network resources but it also depends on other concurrent demands. For a multiple-flow network the residual capacity of a link is calculated as:
\begin{flalign}
r(u,v) = c(u,v) - f(u,v)
\label{formula:res-cap}
\end{flalign} \\  

\noindent in which $c(u,v)$ indicates the (initial) capacity of the link between nodes $u$ and $v$, and $f(u,v)$ indicates the total flow on this link as:\\
\begin{flalign}
f(u,v) = \sum_{k = 1..K}{f_k (u,v)x_k(u,v) }
\label{formula:flows}
\end{flalign} \\

\noindent where $f_k(u,v)$ is the flow number $k$, assuming $K$ concurrent flows. The $x_k(u,v) \in \{0,1\}$ is the entry of Link-Flow-Membership array in the $k$th column and the row corresponding to link $(u,v)$ that determines if the flow number $k$ is passing through link $(u,v)$ or not.

\begin{algorithm}[!b]
\caption{The Residual Capacity predicate, to calculate the residual capacity of all links of a network graph}
\label{alg:Res-Cap-Predicate}

\begin{small}
${1: \textbf{forall}(i\ \textbf{in}\ 1..N_{links})(}$\newline
${2: \ \ \textbf{if}\ \textbf{sum}(j\ \textbf{in}\ 1..N_{flows})(LFM[i,j])=0 }$\newline
3: \ \ \ \textbf{then}\ ${Residuals[i] = MAX}$\newline
${4: \ \ \textbf{else}\ Residuals[i] =}$\newline
${5: \ \ Links[i,3] - \textbf{sum}(j\ \textbf{in}\ 1..N_{flows})}$
${(Flows[j] \times LFM[i,j])}$\newline
6: \ \ \ \textbf{endif}\newline
${\ \ \ \ \ \ \ \ \ \ \ \ \ \ \ \ \ \ \ \ \ \ \ \ \ \ \ \ \ \ )}$\newline
\end{small}

\end{algorithm}

The MiniZinc code for this predicate is shown in Algorithm \ref{alg:Res-Cap-Predicate}. Although the $LFM$ is the main decision variable in the applications implemented in SCOR for routing purposes, $Residuals$, is the decision variable in this predicate. The predicate investigates all links (line 1). When there is no flow on a link (line 2), the residual capacity of the link is set to \emph{MAX} (line 3), a big constant number. This is to prevent some issues that might happen in using the predicate in non-uniform network topologies (i.e. links have different capacities). An example is when an algorithm uses the minimum residual capacity of all links. Suppose there is a link with no flow on it and its capacity is less than the minimum residual capacity of other links (links with flows). In this case, such an algorithm chooses the smaller value, i.e. the capacity of the link with no flow, as the minimum residual capacity. It means that the output of such an algorithm includes a link which is not on any path. Then, using a value (bigger than all capacities), for the residual capacity of such links solves the problem. For the links with flows, the residual capacity is calculated through Eq. \ref{formula:res-cap} and \ref{formula:flows} in lines 4 and 5.\\

\subsubsection{Link Capacity Constraint Predicate}
This predicate enforces the flow bound constraint as stated below

\begin{flalign}
f(u,v) \leq c(u,v)\ \ \ \ \ \ \forall (u,v) \in \mathcal A\
\label{formula:flow_bound}
\end{flalign}\\
\noindent on each link per each flow. It causes the flows to be placed only on links that can accommodate their demands, i.e. ${r(u,v) \geq 0}$ (compare Eq. \ref{formula:res-cap} and Eq. \ref{formula:flow_bound}). While the Path Capacity Constraint predicates serves as an initial stage for applying other constraints, this predicate is used in line with other constraints to guarantee the final output satisfies the flow demands. The MiniZinc code for this predicate is shown in Algorithm \ref{alg:link-constraint-Predicate}. The decision variables in this predicate include the $LFM$ and $Residuals$. The first line is an example of using a predicate in a MiniZinc code. In fact, this is a \emph{nested predicate} which uses another predicate, residual capacity. In the second line it calculates $Residuals$, the residual capacity of the links and the next lines enforce the flow placements to satisfy the positive residual capacity $(Residuals[i] >0)$ for all the links. It is possible to modify this predicate in order to implement  routing with \emph{capacity overbooking}.\\

\begin{algorithm}[!b]
\caption{The Link Capacity Constraint predicate, to enforce link constraint}
\label{alg:link-constraint-Predicate}

\begin{small}
${1: \textbf{include}\ "Predicate\_residual\_capacity.mzn";}$\newline
${2: residual\_capacity( LFM, Flows, Links, Residuals ) \bigwedge}$\newline
${3: \textbf{forall}(i\ \textbf{in}\ 1..N_{links})}$\newline
${4: \ \ (Residuals[i]>0)}$
\end{small}

\end{algorithm}

\subsubsection{Path Cost Predicate}      
This is the last predicate of the first group of predicates that are used to model basic and composite QoS routing problems. It calculates the total cost of a  path $P$ for the $k$th flow given by:

\begin{flalign}
a_k(P) = \sum_{(u,v) \in Nodes}{a_k (u,v) x_k (u,v)}
\label{formula:costs}
\end{flalign} 
\begin{center}
${x_k (u,v) \in \{0,1\} }$
\end{center}

\noindent in which $a_k (u,v)$ is the cost of the link $(u,v)$ for the $k$th flow. The cost is assumed a known predefined parameter such as loss ratio, link or bandwidth price, etc., that can be different per flow or be the same. The MiniZinc code for this predicate is shown in Algorithm \ref{alg:Total-cost-Predicate}. The first decision variable in this predicate is $LFM$ which is used to locate flows and the next decision variable is ${Total\_Cost}$ which implements Eq. \ref{formula:costs} to compute the total cost of a flow path (lines 2, 3 and 4).\\  
   
\begin{algorithm}[!t]
\caption{Path Cost predicate, to calculate the the total cost of a path}
\label{alg:Total-cost-Predicate}

\begin{small}
${1: \textbf{forall}(j\ \textbf{in}\ 1..N_{flows})(}$\newline
${2: \ \ Total\_Cost[j] = }$\newline
${3: \ \ \textbf{sum}(i\ \textbf{in}\ (1..N_{links})}$\newline
${4: \ \ \ \ (Links[i,4] \times LFM[i,j])}$\newline
${\ \ \ \ \ \ \ \ \ \ \ \ \ \ \ \ \ \ \ \ \ \ \ \ \ \ \ \ \ \ \ )}$\newline
${\ \ \ \ \ \ \ \ \ \ \ \ )}$
\end{small}

\end{algorithm}

\subsubsection{Delay Predicates}
The Delay, Congestion and Link Utilisation belong to the second group of predicates. While the first group uses exact measures for the implemented parameters, the second group uses approximations to find the value of these parameters. This is due mainly to lack of the necessary information for the calculation of accurate values. In all predicates of this group, it is assumed that the network flow behaviour can be modelled by a M/M/1 process. This is an assumption made and utilised in highly cited references such as \cite{Data_networks} and \cite{Netwrok_Flows_Theory_Algorithms_and_Applications}. In queueing theory, the M/M/1 systems consists of a single queueing station with a single server, requests arrive according to a Poisson process, and the server passes customers in the exponentially-distributed independent times. Adopting M/M/1 system for modelling network flow behaviours require to accept the Poisson distribution for the network flows which might not be the case in some occasions. Though, other assumptions about the network flow statistics that may result in different equations can be implemented in SCOR in a similar manner.

The end-to-end delay of a flow path is the sum of the delays experienced by the flow at each node from the source to destination. The delay at each node  consists of a fixed and a variable component. The fixed component includes the transmission and propagation delays, and the variable component includes the processing and queueing delays at each node \cite{End-to-end_packet-delay-and_loss-behavior-in_the_internet}. The average variable delay is closely related to the congestion and link utilisation that depend on the flow demand. The fixed part of the end-to-end delay can be calculated by the `\emph{Path Cost}' predicate provided the link cost is defined as the link delay.

\begin{algorithm}[!b]
\caption{The Delay predicate, to calculate the total queueing delay of a flow path}
\label{alg:Delay-Predicate}

\begin{small}
${1: \textbf{include}\ "Predicate\_link\_capacity.mzn";}$\newline
${2: link\_capacity( LFM, Flows, Links, Residuals ) \bigwedge}$\newline
${3: \textbf{forall}(i\ \textbf{in}\ 1..N_{links})(}$\newline
${4: \ \ \textbf{if}\ \textbf{sum}(j\ \textbf{in}\ 1..N_{flows})(LFM[i,j])=0 }$\newline
5: \ \ \ \textbf{then}\  ${Delay[i] = 0}$\newline
${6: \ \ \textbf{else}\ Delay[i] = 1\ /\ Residuals[i]}$\newline
${7: \ \ \textbf{endif}}$ \newline
${\ \ \ \ \ \ \ \ \ \ \ \ \ \ \ \ \ \ \ \ \ \ \ \ \ \ \ \ \ \ \ )}$
\end{small}

\end{algorithm}

The Delay predicate calculates the total average variable delay of a network topology using the delay formula for M/M/1 systems \cite{Introduction_to_queueing_theory}:

\begin{flalign}
Delay = \sum_{(u,v)\in Nodes}{\frac{1}{c(u,v) - f(u,v)}}
\label{formula:Delay}
\end{flalign}  \\

\noindent in which $c(u,v)$ is the link capacity and $f(u,v)$ is the total flow passing through the link. The MiniZinc code for this predicate is shown in Algorithm \ref{alg:Delay-Predicate}. This is another example of a nested predicate in which the Link Capacity constraint predicate is utilised (line 1). Creating abstractions is an important aspect of the Northbound APIs. While the Link Capacity constraint predicate is a nested predicate that uses the Residual Capacity predicate, the Delay predicate creates another level of abstraction by using it. Although the decision variables in this predicate include the $Delay$, $LFM$ and $Residuals$ arrays, the main variable is $Delay$ and the value of other variables are assumed known. The second line enforces the link capacity constraint which prevents from link saturation and calculates the residual capacity of the links. Then all links are investigated (line 3) to see if there is any flow on the links (line 4). The value of queueing delay for the flow-less links (links without any flow) is set to zero (line 5) and for the links which include flows it is the inverse of the remaining capacity of the link (line 6).\\

\subsubsection{Congestion Predicates}
The Congestion predicate calculates the the total network congestion through \cite{Netwrok_Flows_Theory_Algorithms_and_Applications}:

\begin{flalign}
Congestion = \sum_{(u,v)\in Nodes}{\frac{f(u,v)}{c(u,v) - f(u,v)}}
\label{formula:congestion}
\end{flalign}\\

\noindent The MiniZinc code for this predicate is shown in Algorithm \ref{alg:Congestion-Predicate}. The implementation is quite similar to the Delay predicate with a difference that uses Eq. \ref{formula:congestion} instead of Eq. \ref{formula:Delay}. The decision variables in this predicate include ${Congestion}$, $LFM$ and ${Residuals}$ arrays and the ${Congestion}$ is the main variable.\\

\begin{algorithm}[!b]
\caption{The Congestion predicate, to calculate the total congestion of a flow path}
\label{alg:Congestion-Predicate}

\begin{small}
${1: \textbf{include}\ "Predicate\_link\_capacity.mzn";}$\newline
${2: link\_capacity( LFM, Flows, Links, Residuals ) \bigwedge}$\newline
${3: \textbf{forall}(i\ \textbf{in}\ 1..N_{links})(}$\newline
${4: \ \ Congestion[i] = }$\newline
${5: \ \ \textbf{sum}(j\ \textbf{in}\ 1..N_{flows})}$\newline
${6: \ \ \ \ \ (Flows[j] \times LFM[i,j])\ / \ Residuals[i]}$\newline
${\ \ \ \ \ \ \ \ \ \ \ \ \ \ \ \ \ \ \ \ \ \ \ \ \ \ \ \ \ \ \ )}$
\end{small}

\end{algorithm}

\subsubsection{Link Utilisation Predicates}
This predicate uses the below equation \cite{Routing_flow_and_capacity_design_in_communication_and_computer_networks}:

\begin{flalign}
Utilisation = \sum_{(u,v)\in Nodes}{\frac{f(u,v)}{c(u,v)}}
\label{formula:utilisation}
\end{flalign}\\

\begin{algorithm}[!t]
\caption{The Link Utilisation predicate, to calculate the link utilisation of a flow path}
\label{alg:Link-Utilisation-Predicate}

\begin{small}
${1: \textbf{forall}(i\ \textbf{in}\ 1..N_{links})(}$\newline
${2: \ \ Link\_Utilisation[i] = }$\newline
${3: \ \ \textbf{sum}(j\ \textbf{in}\ 1..N_{flows})(}$\newline
${4: \ \ \ \ \ \ Flows[j] LFM[i,j])/\ Link[i,3]) \times 100}$\newline
${\ \ \ \ \ \ \ \ \ \ \ \ \ \ \ \ \ \ \ \ \ \ \ \ \ \ \ \ \ \ \ \ )}$\newline
${\ \ \ \ \ \ \ \ \ \ \ \ \ \ \ \ \ \ \ \ \ \ \ \ \ \ \ \ \ \ )}$
\end{small}

\end{algorithm}

\noindent to calculate link utilisation. The MiniZinc code for this predicate is shown in Algorithm \ref{alg:Link-Utilisation-Predicate}. It investigates all links (line 1) and calculates the link utilisation (line 2,3 and 4) based on the Eq. \ref{formula:utilisation}. The decision variables in this predicate include the $Link\_Utilisation$ and $LFM$ arrays, with $Link\_Utilisation$ as the main variable and the $LFM$ which is seen as a known value.\\

\section{Use cases: Implemented Applications}
The main propose of SCOR is to facilitate QoS routing application development in SDN which is not currently addressed in the available northbound interfaces. By using a few examples, it is shown in this section that, how easily, simple and complex QoS routing and TE applications are implemented in SCOR. Each QoS routing application is implemented by including the Network Path predicate to model the network paths, and some other predicates to apply their objective constraints on the paths. The first application is the least cost path routing. The next application is least cost path routing with the capacity constraint which is simply implemented by adding a new predicate to the previous application. This is an example of the abstractions created in the SCOR and its accommodation for developing QoS routing applications. The last application is a complex TE application which is simply modelled in SCOR. This clearly illustrates SCOR potential for this purpose. The routing algorithms modelled in these applications are explained briefly in this section and might require more details and discussion, the interested reader can refer to the references for further information.\\

\subsection{Least cost Path Routing}
The first use case models the least cost path routing algorithm. It is single model that can model multiple routing algorithms based on the definition of cost. If the link delay is defined as the cost it will give the minimum delay path (fixed part of delay) and if the distance (hop count) is used as the cost metric, it will give the shortest path. The shortest path routing is the most common network routing algorithm in legacy networks which is used in many routing protocols such as OSPF, RIP and IGRP. 

The formal statement of the problem is presented in Section \ref{Background} (Eq. \ref{formula:shortest_path_1}). Several algorithms such as Dijkstra and Bellman-Ford have been proposed to efficiently solve this problem in procedural programming. In SCOR though, the implementation only includes stating the problem, and not how to solve it, as it can be seen in Algorithm \ref{alg:SP-MiniZinc}.  The first two lines (line 2 and 3, lines starting with a {\scriptsize $\%$} are comments) are \emph{include items} which make it possible to use two predicates, Network Path and Path Cost. Lines 5-11 declare the \emph{parameters} which are required to model a least cost path problem (same as Table \ref{tab:Predicate_Names}). Lines 13-14 also declare the two decision variables $Cost$ and $LFM$ similar to what explained in the SCOR predicates. The main body of the program includes the \emph{Constraints item} (lines 16-17) and the \emph{Solve item} (line 19). The first constraint, Network Path predicate, lists all possible paths from the source to destination and the second constraint, Path Cost predicate, calculates the cost associated with each path. The \emph{solve item} indicates the objective, i.e. what the solver should find that is the path has the minimum cost.\\

\begin{algorithm}[!b]
\caption{Least Cost Path in SCOR}
\label{alg:SP-MiniZinc}

\begin{flushleft}
\begin{small}
${1: \ \%\ Include\ item}$\newline
${2: \  \textbf{include}\ "Predicate\_network\_path.mzn";}$\newline
${3: \ \textbf{include}\ "Predicate\_path\_cost.mzn";}$\newline

${4: \ \%\ Parameters}$\newline
${5: \ \textbf{array}[int, int]\ of\  \textbf{int}:\ Links; }$\newline
${6: \ \textbf{int}:\ N_{links}\ =  \textbf{max}(\textbf{index\_set\_1of2}(Links));}$\newline
${7: \ \textbf{array}[int]\ of\  \textbf{ int}:\ Nodes;}$\newline
${8: \ \textbf{array}[int]\ of\  \textbf{int}:\ Flows;}$\newline
${9: \ \textbf{int}:\ N_{flows}\ =\  \textbf{max}(\textbf{index\_set}(Flows));}$\newline
${10: \textbf{array}[1..N_{flows}]\ of\  \textbf{int}:\ s;}$\newline
${11:  \textbf{array}[1..N_{flows}]\ of\  \textbf{int}:\ d;}$\newline

${12: \%\ Decision\ Variables}$\newline
${13:  \textbf{array}[1..N_{flows}]\ of\  \textbf{var}\ \textbf{int}:\ Cost;}$\newline
${14: \textbf{array}[1..N_{links},1..N_{flows}]\ of\  \textbf{var}\ 0..1:\ LFM;}$\newline

${15: \%\ Constraints\ item}$ \newline
${16: \textbf{constraint}\ network\_path( LFM, Links,Nodes, s, d);}$
${17:  \textbf{constraint}\ path\_cost( LFM, Links, Cost);}$\newline

${18: \% Solve\ item}$\newline
${19: \textbf{solve}\  \textbf{minimize}\ Cost[1];}$
\end{small}

\end{flushleft}
\end{algorithm}


\subsection{Least Cost Path Routing with Capacity Constraint}
In order to illustrate the usefulness of SCOR, another implemented application is demonstrated here. This application uses all code of the previous application plus a single line of code which implements a new constraint on network parameters. This is a clear example of combining various pre-implemented predicates in SCOR to create numerous applications with multiple constraints. 

The path selected by the least cost path routing algorithm does not necessarily provide sufficient bandwidth for a flow demand. If the selected path's capacity or bandwidth is less than the flow demand it can cause delay or even packet loss. When the service level is guaranteed for a flow demand it is necessary to prevent this by applying QoS mechanisms i.e. considering the capacity constraint when finding the path. The formal statement of the problem is similar to least cost path routing, except there is an extra constraint, Eq. \ref{formula:flow_bound} the capacity constraint, which should be satisfied. 

As seen in Algorithm \ref{alg:SPC-MiniZinc} it is implemented in SCOR by adding the three lines of code to the previous application. The first added line (line 2) indicates a new predicate, Capacity Constraint, is used in this application and the second added line (line 4) declares a new parameter, $Limits$ (explained in Table \ref{tab:Predicate_Names}). Finally, the third added line (line 6) applies the capacity constraint. This constraint removes the links with capacities less than $Limits$ from the network graph. Consequently the output of the algorithm not only is the path with least cost from `${s}$' to `${d}$' but it can also accommodate the flow demands to the specified $Limits$.\\

\begin{algorithm}[!b]
\caption{Least Cost Path with Capacity Constraint in SCOR (only lines not in Least Cost Path predicate)}
\label{alg:SPC-MiniZinc}

\begin{flushleft}
\begin{small}
${}$\newline
.\newline
${1: \ \%\ Include\ item}$\newline
${2: \textbf{include}\ "Predicate\_capacity\_constraint.mzn";}$\newline
.\newline
.\newline
${3: \ \%\ Parameters}$\newline
${4:  \textbf{array}[1..N_{flows}]\ of\  \textbf{int}:\ Limits;}$\newline
.\newline
.\newline
${5: \%\ Constraints\ item}$ \newline
${6: \textbf{constraint}\ capacity\_constraint( LFM, Flows,Links, }$\newline
${. \ \ \ \ \ \ \ \ \ \ \ \ \ \ \ \ \ \ \ \ \ \ \ \ \ \ \ \ \ \ \ \ \ \ \ \ \ \ \ \ \ \ \ \ \ \ \ \ \ \ \ \ \ \ \ \ \ \ \ \ \ \ \ \ \ \ Limits);}$\newline
.
\end{small}
\end{flushleft}

\end{algorithm}


\subsection{Maximum Residual Capacity Routing}
The previous application illustrates the efficacy of abstractions and modularity in SCOR. It not only shows illustrates the easiness of creating new applications by adding a new predicates or combining them, but it also demonstrate the compliance of stating the problem in SCCOR and its statement in natural language. Here, implementation of a complex QoS routing application in SCOR is compared to its implementation in procedural programming. This demonstrates how efficiently SCOR simplifies the QoS routing and TE application development.

The \emph{Maximum Residual Capacity} belongs to a group of routing algorithms that aim to route the maximum number of concurrent flows in a given network. In these algorithms, usually the given flows must be routed between single or multiple source-destination pairs and then, algorithms try to either  minimise the congestion or maximise the utilisation. In \emph{Maximum Residual Capacity} the objective is to route all flows in a way that the minimum residual capacity of the links is maximised \cite{Netwrok_Flows_Theory_Algorithms_and_Applications}. This is a very complex network routing problem which can be numerically intractable even for small networks of a few nodes \cite{P-Maximizing_residual_capacity_in_connection_oriented_networks}.

\begin{algorithm}[!b]
\caption{Maximum Residual Capacity in SCOR}
\label{alg:MRC-MiniZinc}

\begin{flushleft}
\begin{small}
${1: \ \%\ Include\ item}$\newline
${2: \ \textbf{include}\ "Predicate\_network\_path.mzn";}$\newline
${3: \ \textbf{include}\ "Predicate\_residual\_capacity.mzn";}$\newline

${4: \ \%\ Parameters}$\newline
${5: \ \textbf{array}[int, int]\ of\  \textbf{int}:\ Links; }$\newline
${6: \ \textbf{int}:\ N_{links}\ =  \textbf{max}(\textbf{index\_set\_1of2}(Links));}$\newline
${7: \ \textbf{array}[int]\ of\  \textbf{ int}:\ Nodes;}$\newline
${8: \ \textbf{array}[int]\ of\  \textbf{int}:\ Flows;}$\newline
${9: \ \textbf{int}:\ N_{flows}\ =\  \textbf{max}(\textbf{index\_set}(Flows));}$\newline
${10: \textbf{array}[1..N_{flows}]\ of\  \textbf{int}:\ s;}$\newline
${11: \textbf{array}[1..N_{flows}]\ of\  \textbf{int}:\ d;}$\newline

${12: \%\ Decision\ Variables}$\newline
${13: \textbf{array}[1..N_{links},1..N_{flows}]\ of\  \textbf{var}\ 0..1:\ LFM;}$\newline
${14: \textbf{array}[1..N_{links}]\ of\  \textbf{var}\ \textbf{int}:\ Residuals;}$\newline

${15: \%\ Constraints\ item}$ \newline
${16: \textbf{constraint}\ network\_path( LFM, Links,Nodes, s, d);}$
${17: \textbf{constraint}\ residual\_capacity( LFM, Flows,Links,}$\newline
${\ \ \ \ \ \ \ \ \ \ \ \ \ \ \ \ \ \ \ \ \ \ \ \ \ \ \ \ \ \ \ \ \ \ \ \ \ \ \ \ \ \ \ \ \ \ \ \ \ \ \ \ \ \ \ \ \ \ \ \ \ \ Residuals);}$\newline
${18: \textbf{constraint}\ \textbf{forall}(i\ \textbf{ in}\ 1..N_{links})(Residuals[i] >= 0);}$\newline

${19: \% Solve\ item}$\newline
${20: \textbf{solve}\ \textbf{maximize}\ \textbf{min}(Residuals);}$
\end{small}

\end{flushleft}
\end{algorithm}


For a given graph ${\mathcal G(\mathcal N,\mathcal A)}$ and a set of flows $f$, the problem can be mathematically stated as \cite{P-Maximizing_residual_capacity_in_connection_oriented_networks}:
\[\textsl{maximize}\ \ \ \ \ Z \]
subject to
\begin{flalign}
\label{formula:mrc-1}
Z \leq r(u,v)\ \ \ \ \ \ \ \ \ \ \ \forall (u,v) \in \mathcal{A}
\end{flalign}

\begin{flalign}
\label{formula:mrc-2}
r(u,v) = c(u,v) - f(u,v)
\end{flalign}

\begin{flalign}
\label{formula:mrc-3}
\sum_{\{v|(u,v) \in  \mathcal A\}}{f(u,v)} - \sum_{\{v|(u,v) \in  \mathcal A\}}{f(v,u)}= 
	\begin{cases}
		\ \ 1\ \ if\ u=s,\\
		-1 \ \ if\ u=t,\\
		\ \ 0\ \textsl{otherwise}
	\end{cases}
\end{flalign}

\begin{flalign}
\label{formula:mrc-4}
f(u,v) = \sum_{k = 1..K}{f_k (u,v)x_k(u,v) }
\end{flalign}\\

\noindent where Eq. \ref{formula:mrc-2}, \ref{formula:mrc-3} and \ref{formula:mrc-4} are same as Eq. \ref{formula:res-cap}, \ref{formula:flow_conservation} and \ref{formula:flows} respectively. There is an implementation of the maximum concurrent flow routing in \cite{Python_for_Maximum_concurrent_flow_solver} that can be adopted to solve this problem. It is implemented in 400 lines of code in Python language based on the solution proposed in \cite{Faster_and_simpler_algorithms_for_multicommodity_flow_and_other_fractional_packing_problems}. The implementation of the \emph{Maximum Residual Capacity} algorithm in SCOR is shown in Algorithm \ref{alg:MRC-MiniZinc}. Lines  2-3 are \emph{include item} for the Network Path and Residual Capacity predicates. Lines 5-11 declare \emph{parameters}
and lines 13-14 declare the two \emph{decision variables}, $Residuals$ and $LFM$.

The main body of the program consists of the three constraints (lines 16-18) and the \emph{solve item} (line 20). The \emph{Network Path predicate} lists paths for all the source-destination pairs, the \emph{Residual Capacity predicate} calculates the residual capacity in each path, and the last constraint enforces the flow placements to satisfy $(Residuals[i] \geq 0)$. The \emph{solve item} indicates that the solver should select a path for each flow that maximises the minimum residual capacity of the whole network.

\section{Evaluation and Results}\label{Evaluation and Results}
The previous section illustrates the usefulness of SCOR by demonstrating the efficacy of its abstractions and explaining the simplicity of developing complex applications in SCOR. In order to evaluate these applications, two other aspects must be also investigated, the completeness and the scalability. The first question is already answered in Section \ref{Design} where the design of predicates is discussed. It is shown there that the set of eight implemented predicates in SCOR can model most of the basic and composite QoS routing algorithms that deal with capacity, delay, jitter etc. Furthermore, the last three predicates make it possible to model advanced TE problems, such maximum concurrent flow problem. In order to answer the second question, it is needed to investigate SOCR in terms of applications implemented in it, i.e verify if they operate within a reasonable time frame. In fact, the suitability of SCOR as a Northbound is validated through evaluating the run time of these applications. \\

\subsection{Evaluation method}
Numerical experiments are necessary in order to evaluate these applications in real networks dimension. There are several parameters that might affect the run time of these applications on a real network. The three parameters, network topology, network size and the number of concurrent flows are considered in designing these experiments. Though, the list of effective parameters may not be limited to this and one can investigate other parameters, e.g. the effect of various solvers, which is an ongoing research topic in CP. \\

\subsubsection{Network Topology}
\begin{figure}[!t]
\centering
\includegraphics [width=7.5cm, height=6cm, natwidth=500, natheight=540]{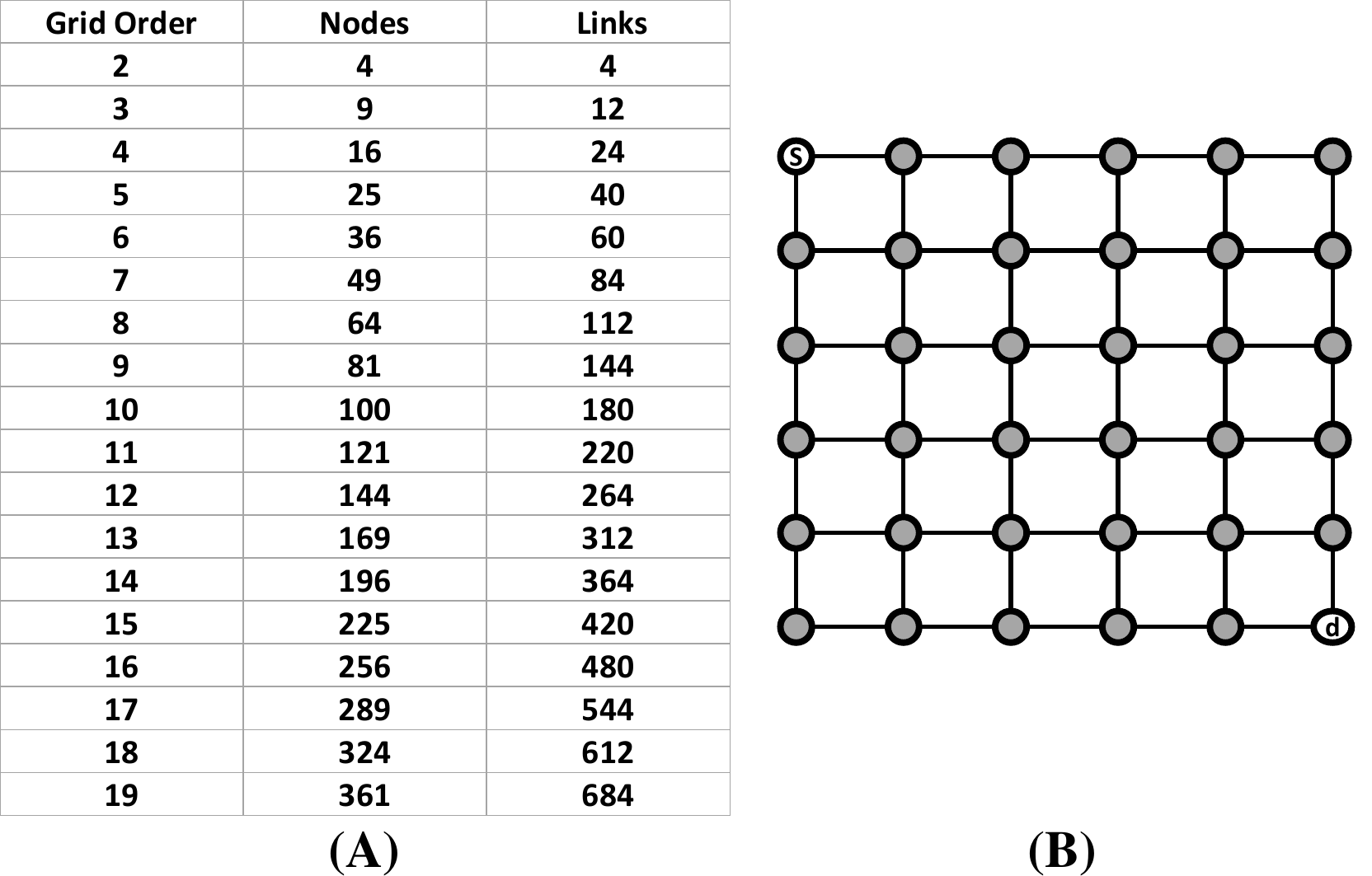}
\caption{A) Network sizes used for the grid topologies. B) A sample grid topology of order 6}
\label{fig:grid}
\end{figure}

Network topology is the trivial effective parameter in the evaluation of routing algorithms. Topologies used in this study include \emph{grid} and \emph{fat-tree} topologies. The grid or mesh topology is common in computer networks such as sensor networks. Since it provides a variety of paths between source-destination pairs, which might be challenging for finding the appropriate path, it is selected for the run-time evaluation of the implemented applications. A sample grid topology is shown in Fig. \ref{fig:grid}-B. In this topology each node is connected to all its neighbours through four links in the case of internal nodes and two or three links for the border nodes. In all experiments in this section, the flow sources and sinks (destinations) are placed at the ends of the grid's diagonal.

\begin{figure}[!b]
\centering
\includegraphics [width=7.5cm, height=6cm, natwidth=750, natheight=650]{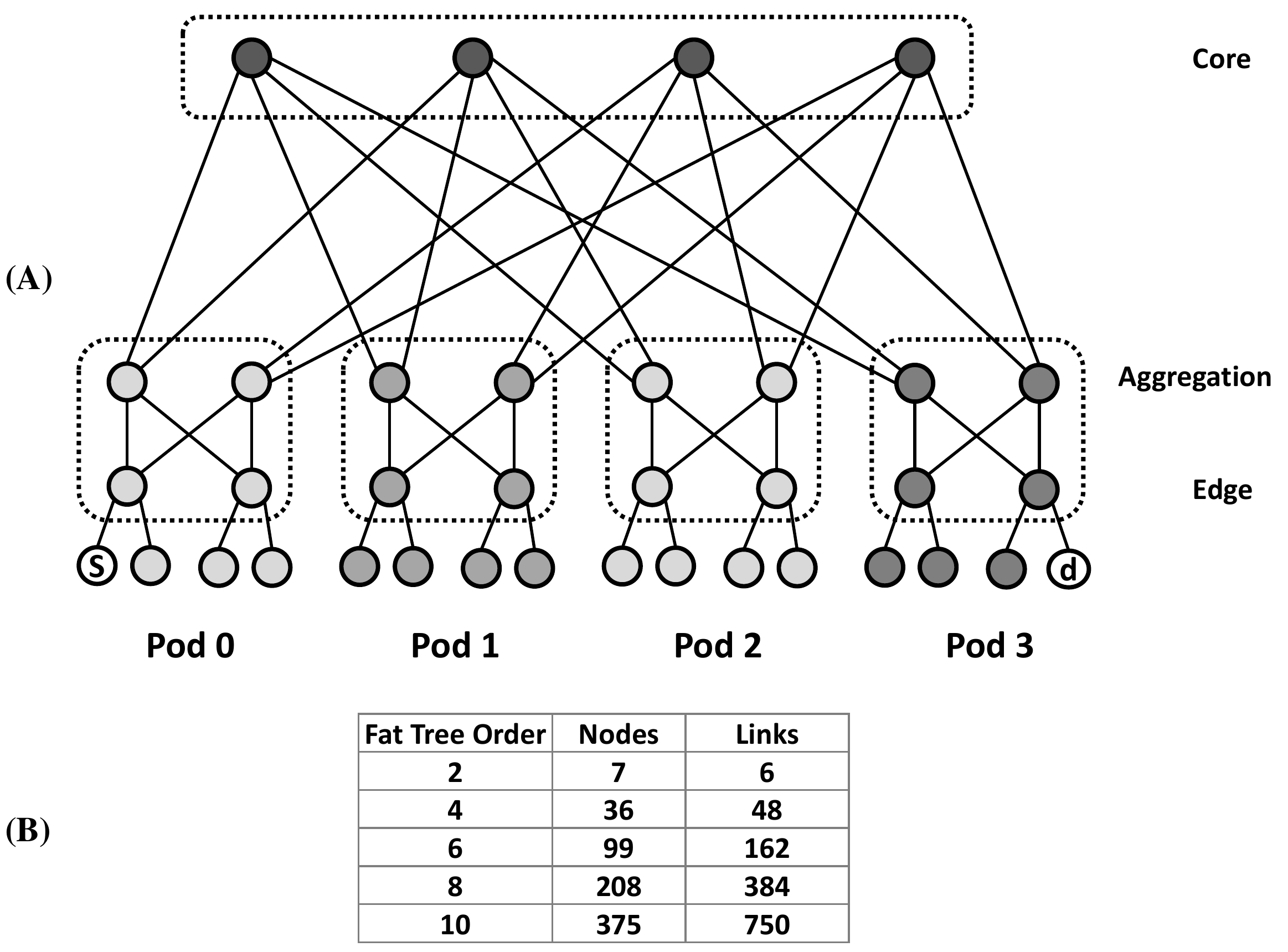}
\caption{A) The adopted fat tree topology for SDN switches of order k=4 \cite{P-A_scalable_commodity_data_center_network_architecture} B) Network sizes used for the fat tree topologies.}
\label{fig:fat_tree}
\end{figure}

The fat-tree topology used in this evaluation is suggested by \cite{P-A_scalable_commodity_data_center_network_architecture} for interconnecting the SDN switches in data centres which is a special instance of the fat-tree topology. In the main fat-tree topology each layer is connected to the higher layer through a single link with higher capacity (rather than multiple equal-capacity links). This necessitates to use high capacity switching devices for the core and aggregation layers. Though, Since all the nodes are identical in the adopted fat-tree topology, it is possible to use the cheap commodity switches in the core and aggregation layers. A sample adopted fat tree topology is depicted in Fig. \ref{fig:fat_tree}-A. The main idea is to make a topology of similar switches with equal number of ports ($k$). The fat tree order $k$ includes $k$ pods, containing $k$ switches in two layers, aggregation and edge layers. Below the pods is the place where the hosts are located and each $k/2$ of them are directly connected to one $k$-port edge switch. The remaining $k/2$ ports of the edge switches are connected to the aggregation switches which in turn their remaining $k/2$ ports are connected to the core switches. Each of the $(k/2)^2$ core switches is connected to all $k$ pods.

\subsubsection{Network Size}
Basically, size of the network determines its situation in real world applications. For instance, a real world carrier network might include hundreds of nodes while a campus network might only include tens of nodes. Indeed, size of the network is a significant scalability factor that can be used as a parameter in the run-time evaluation of the routing algorithms. The network sizes used in this study for the grid and fat tree topologies are shown in Fig. \ref{fig:grid}-A and \ref{fig:fat_tree}-B respectively . \\

\subsubsection{Number of Concurrent Flows}
If flow specifications are considered for the path calculations in a routing algorithm, the number of concurrent flows is another essential parameter in the run-time evaluation. One significant aspect of routing multiple flows, instead of a single flow, is the effect of selecting order. In routing a single flow, only constraints relating to the same flow, e.g. the bandwidth to accommodate the flow demand, is considered. In routing multiple concurrent flows though, not only a flow's own constraints is considered, but the various cases for the residual network (Section \ref{Implementation}), that might happen corresponding to each flow, must be considered too. 

Whenever there are multiple flows, the order of placing flows will result in a different residual network. Owing to the fact that there might be various paths for routing a single flow in a network, there might be multiple different residual networks after routing the first flow. If another flow was routed first, other set of residual networks would have been generated for routing the next flows. Consequently, the number of possibilities for choosing path and so the complexity of the problem is increasing exponentially that significantly affects the run time of routing application.\\

\subsection{Results}
The set up used in these experiments includes a virtual machine with 4 virtual CPUs, each with 100\% execution capacity, and 11 GB memory. The virtual machine runs on a host computer with a 64-bit quad core CPU at a clock speed of 3.6 GHz and 16 GB of memory. The host machine runs a 64-bit Windows 7 Enterprise operating system and the guest machine is a Linux with the kernel version of 3.13.0-24-generic. The virtualisation platform is the Oracle VirtualBox hypervisor version 5.0.2 with an extension pack version 5.0.2\_106931. The SCOR is running on version 2.0.2 of G12 MiniZinc package and the implemented applications are evaluated through an interface developed in Python 2.7. The two solvers used in these experiments include Gecode version 4.4.0 and mzn-g12mip (mixed integer programming) which come with MiniZinc version 2.0.2.

The three implemented applications are exactly evaluated according to the above mentioned method. For the effect of topology and size, all the three applications, the Least Cost Path, the Least cost Path with Capacity Constraint and Maximum Residual Capacity routing applications are used in the experiments. Though, for the effect of various number of concurrent flows, only the Maximum Residual Capacity routing application is used.

\begin{figure}[!b]
\centering
\includegraphics [width=7.5cm, height=4.5cm, natwidth=450, natheight=550]{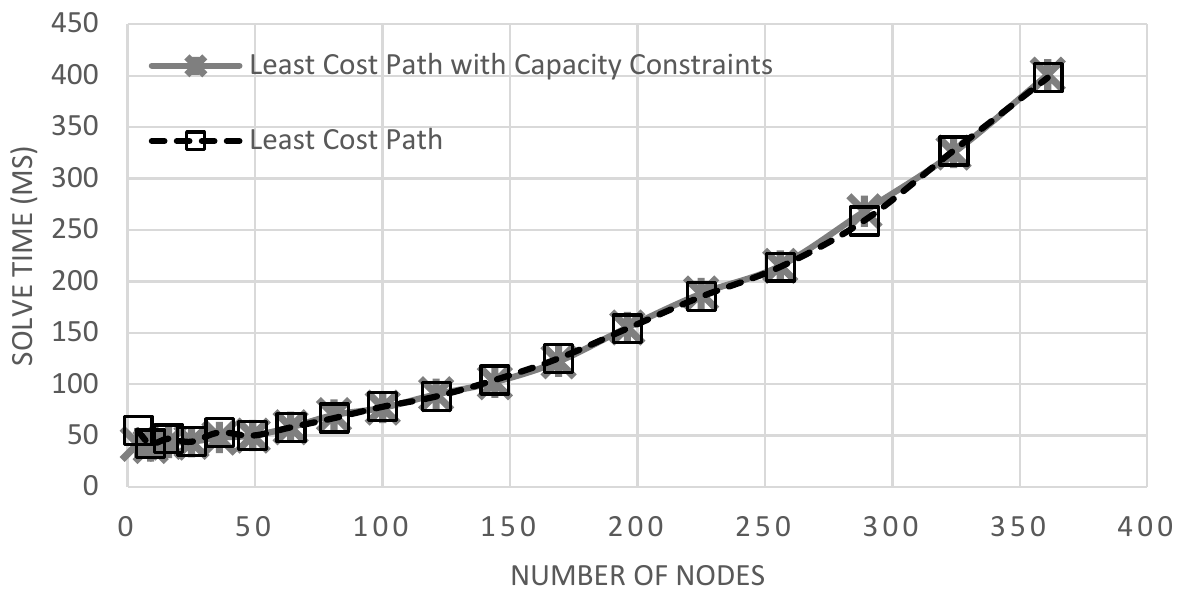}
\caption{The solution time for the Least Cost Path and Least Cost Path routing with Capacity Constraint routing applications in a grid topology of different network sizes (number of nodes) }
\label{fig:LC_LCCC_Grid_MIP}
\end{figure}

There are network emulation environments such as Mininet \cite{Mininet} that are commonly used in SDN to implement various experiments in a single PC. According to \cite{Mininet}, Mininet can be used to implement large network topologies. However, when there are resource-consuming applications, the practical limit is much smaller due to limitations in system resources such as RAM and CPU. In this case, the resources might not suffice for the applications and the run-time evaluation might be inaccurate. A possible solution is to evaluate these applications without the routing framework, by sending the network graphs and flow statistic through a separately developed interface in python. The graphs for the grid and fat-tree topologies of various sizes are created and their adjacency lists are sent to the applications. The graphs in all experiments include links with a unique capacity of 1000 Mbps in both network topologies. The flow demands are also set equal to 10 Mbps in all experiments. For the grid topologies the source and sink nodes are assumed at the end of grid's diagonal (similar to Fig. \ref{fig:grid}-B) for all flows. In the fat-tree topologies the source and sink nodes are the first and last hosts (similar to Fig. \ref{fig:fat_tree}-A) for all flows.
   
Fig. \ref{fig:LC_LCCC_Grid_MIP} displays the solve time of the Least Cost Path and Least Cost Path with Constraint Capacity routing applications for grid networks of various sizes. The horizontal axis indicates the number of nodes which represents the network size the vertical axis is the solve time in millisecond. The solve time comprises, in fact, three parts modelling, i.e. converting the model to a set of finite constraints, solving, and mapping the solution to the original problem. While the term `solve time' in references about CP is mostly used for the second part of these three parts, in this work it is used for the run time. i.e the total time spent to call the routing application from the Python interface until receiving the results. As seen, the run-time / solve-time values for a grid network of the size 350 nodes or more is about 400 ms. This is in the range of the shortest path algorithm using Dijkstra implemented in python. 

\begin{figure}[!t]
\centering
\includegraphics [width=7.5cm, height=4.5cm, natwidth=450, natheight=550]{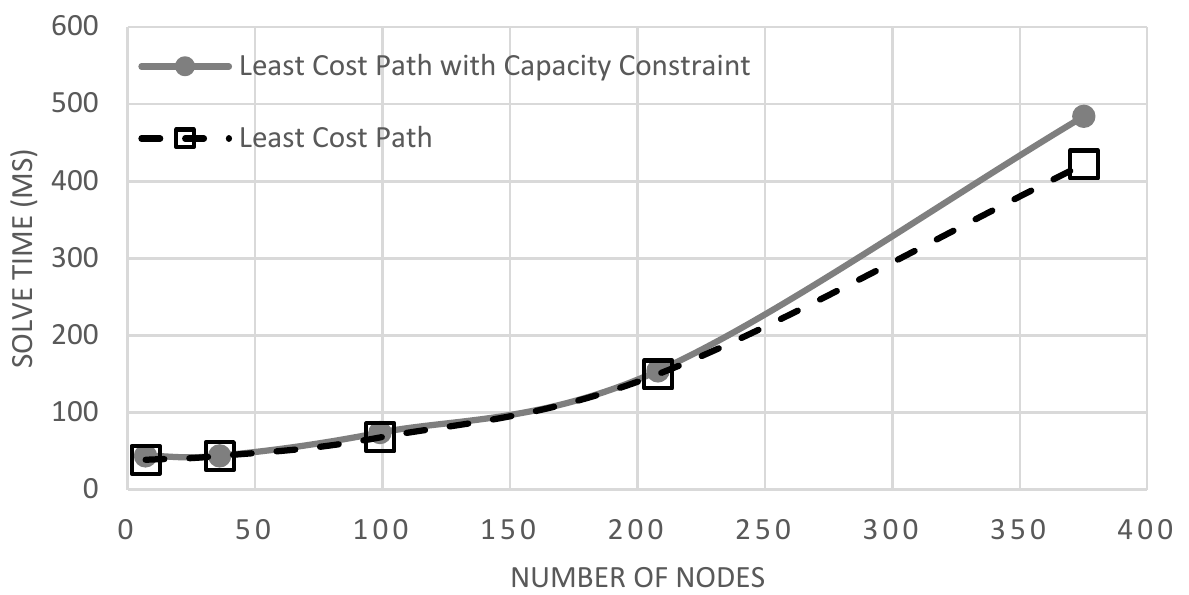}
\caption{The solution time for the Least Cost Path and Least Cost Path routing with Capacity Constraint routing applications in a fat tree topology of different network sizes (number of nodes) }
\label{fig:LC_LCCC_Fat_Tree_MIP}
\end{figure}

Fig. \ref{fig:LC_LCCC_Fat_Tree_MIP} displays the solve time of these applications in networks of fat-tree topology of different sizes. The horizontal and vertical axes are similar to Fig.  \ref{fig:LC_LCCC_Grid_MIP}. The figure indicates that for a fat-tree network with 375 nodes and 750 links (fat-tree order 10, i.e. a network of 10 port switches), the time required to calculate the Least Cost Path route is about 400 ms. This is a practical value for the route calculation in such a big network. The route calculation for the Least Cost Path with Capacity Constraint is less than 500 ms which is very close to the solve time of the Least Cost Path.

Comparing the results of the two applications in grid and fat tree topologies of different sizes indicates that their performance is similar and in some cases the Least Cost Path with Capacity Constraint is a few millisecond faster. Since the unique capacities of all links are assigned big enough to accommodate all flows, the capacity constraint does not result in a shortened network.

Nonetheless the graph of the solve time for both topologies tend to be exponential, the results are still practical for the network sizes indicated in the figures. The illustrated results are generated by applying the routing application on homogeneous networks which is the worse case in terms of route calculation. In a more realistic scenario, networks are not homogeneous and the capacity of the links are most probably different. In this case, the the capacity constraint probably might remove some links from the network which results in smaller network topology  i.e. a smaller search scope for the least cost and shorter solve time. Although using heterogeneous networks might improve the run time, it might bind the run time to the selected capacity values. 

\begin{figure}[!t]
\centering
\includegraphics [width=7.5cm, height=4.5cm, natwidth=450, natheight=550]{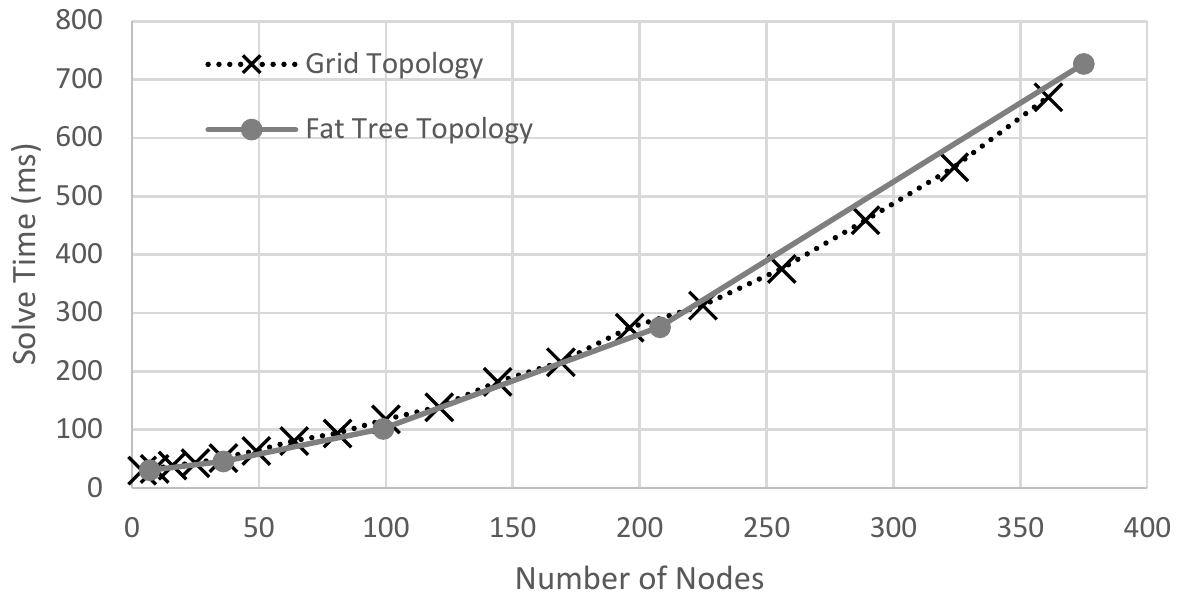}
\caption{The solution time for the Maximum Residual Capacity routing application in the Grid and fat tree topology of different network sizes (number of nodes) }
\label{fig:MRC_Grid_Fat_Tree_1Flow_Gecode}
\end{figure}

The results of Fig. \ref{fig:LC_LCCC_Grid_MIP} and \ref{fig:LC_LCCC_Fat_Tree_MIP} are generated using the mzn-g12mip solver. The other solver which comes with Minizinc in the package, Gecode or in its command line format mzn-gecode, performs much (thousands of milliseconds) slower than mzn-g12mip solver for the above two applications. Though, this is not a general rule for all applications and it is much faster for the next application, the Maximum Residual Capacity routing. This is a clear example of the fact that there is no universally optimised solver for all CP problems, a fact indicated in many references of CP studies \cite{The_Evolving_World_of-MiniZinc}.

\begin{figure}[!b]
\centering
\includegraphics [width=7.5cm, height=4.5cm, natwidth=450, natheight=550]{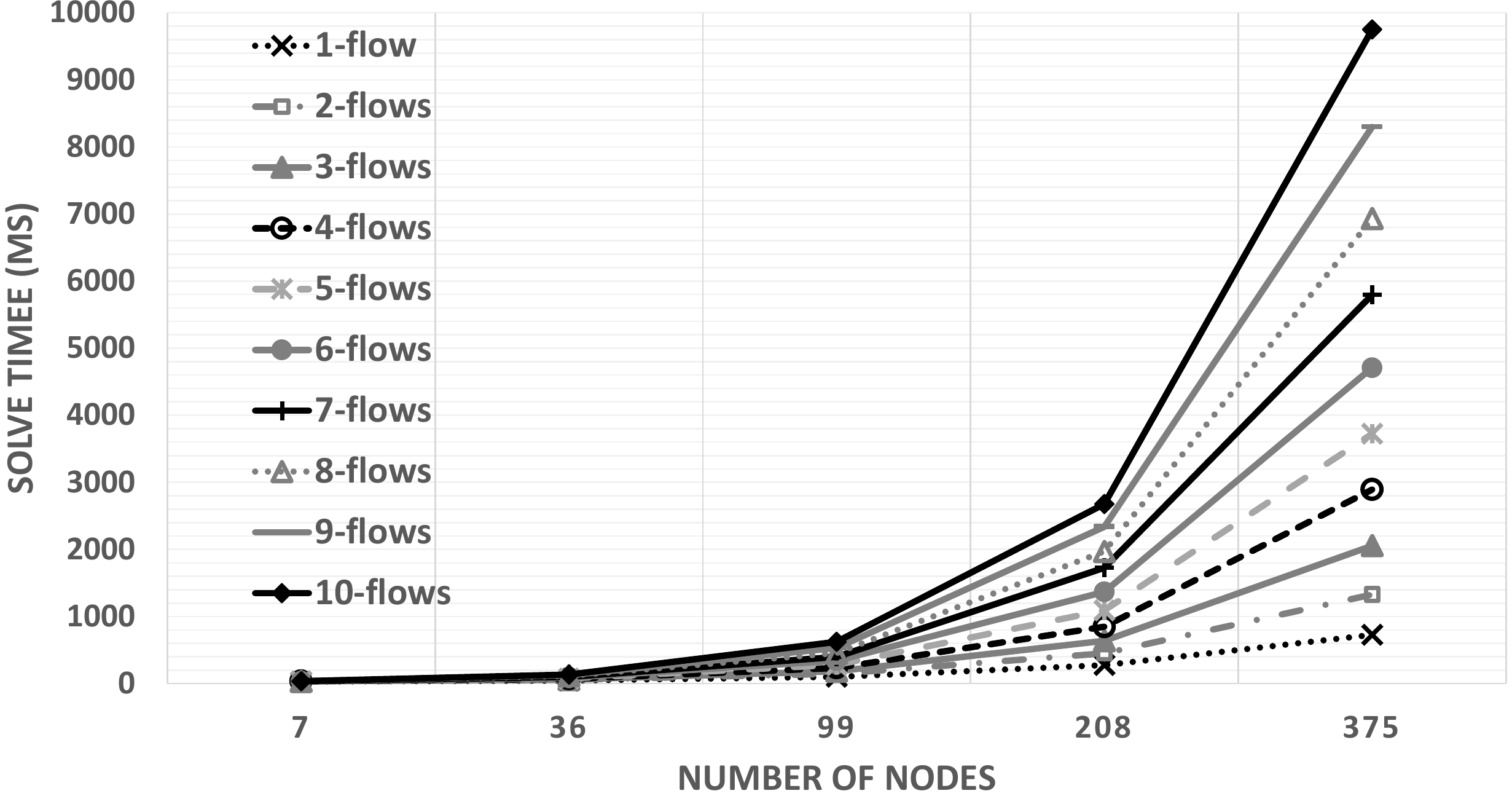}
\caption{The solution time for the Maximum Residual Capacity routing application per various number of concurrent flows in the fat tree topology of different network sizes}
\label{fig:MRC_Fat_Tree_Gecode_Multi_Flow}
\end{figure}

Fig. \ref{fig:MRC_Grid_Fat_Tree_1Flow_Gecode} displays the solve time of the Maximum Residual Capacity routing application on the grid and fat-tree network topologies of various number of nodes using mzn-gecode solver. Again, the horizontal axis indicates the size of the network in terms of number of nodes and the vertical axis is the solve time in milliseconds. It needs about 700 milliseconds to calculate the Maximum Residual Capacity route for a single flow in a network of 375 nodes with the fat tree topology. A similar time (less than 700 milliseconds) is required for the route calculation in a grid network of 361 nodes. For such a complex QoS routing application, which efficiently is implemented in a few lines of code (see Algorithm \ref{alg:MRC-MiniZinc}), the solve times are quite practical in the identified network sizes.

The effect of number of concurrent flows on the run time is illustrated in Fig. \ref{fig:MRC_Fat_Tree_Gecode_Multi_Flow} for the Maximum Residual Capacity routing application. It shows the solve time for various number of concurrent flows in a fat tree topology of various number of nodes. The horizontal axis indicates the network size (number of nodes) and the vertical axis indicates the solve time in milliseconds. As seen, for the small network sizes such as 7 and 36 nodes, adding new flows does not affect the solve time severely. For instance, for 10 concurrent flows the solve time is about 3 times of a single flow. But for the larger network sizes the ratio of the solve time grows faster. For example, for a network size of 375 nodes the solve time of 10 concurrent flows is more than 13 times of a single flow.

Nonetheless, the solve time for 10 concurrent flows is still quite practical. It is 622 ms for a network with 99 nodes and 9750 ms for a network with 375 nodes. In \cite{A_Declarative_and_Expressive_Approach_to_Control_Forwarding_Paths_in_Carrier-Grade_Networks}, a similar TE objective is implemented in linear programming and run on a powerful server which has taken more than few hours for a network of 100 nodes. The SCOR's success is a result of the fact that it uses CP and its intelligent search techniques to solve the problem and find a path among the huge number of possible paths in such network sizes. These results indicate that SCOR not only simplifies developing QoS routing and TE applications in SDN, but it also provides scalable and fully practical applications.\\

\section{Conclusion}\label{Conclusion}

The main contribution of this work is to introduce a new Northbound API for SDN, based on constraint programming techniques. The fundamental goal in the design of this Northbound API is to simplify developing QoS routing and TE applications in SDN. The proposed solution, SCOR, provides building blocks, the predicates, that can be re-used, not only as nested predicates, but also for multiple objectives by combining with other blocks. Applications created in SCOR are quite scalable and practical in large network sizes and can compute excellent solutions to complex TE problems in few second. 

SCOR is implemented in MiniZinc CP modelling language with an extra level of abstraction which forms an interface for QoS routing and TE. The main advantage of using CP techniques in this Northbound interface is to unlock QoS routing application development from complex mathematical / heuristic solution algorithm implementations. In this way, an efficient tool is created that allows to painlessly implement simple and complex QoS routing and TE applications in a few lines of code.

As another contribution, a routing framework is introduced that uses SCOR for its Northbound API. It includes a Network Operating System (NOS) layer that collects network states, such as topologies, capacities, traffic loads and flow demands from the network elements and provides it for upper layers through a network state database. 
The application layer of this routing framework includes various routing applications created through SCOR. They use the abstractions and network states supplied by the NOS to efficiently implement assorted QoS routing and TE applications.

The scalability of the developed applications is evaluated through a series of experiments. The results indicate for a single flow routing problem, either simple or complex, the solve time is quite practical even for real world network sizes. For the case of multiple concurrent flows and complex TE problems such as maximum residual capacity routing, procedural programming techniques such as linear programming might be quite impractical. Though, SCOR provides fully scalable and practical applications that are faster by three orders of magnitude. 

Nonetheless, using CP has provided even more options to improve the scalability. Most of the solvers make it possible for the user to limit the solve time. In this case, the solver will give the best solution that is found in the designated time frame. The results of experiments indicate that the solutions achieved even in one tenth of a full solve-time are very similar to the final solution for practical purposes. This feature can be utilised in the cases wherever tiny non-optimality can be justified by faster solve times.

SCOR not only makes it possible to model and implement different available QoS routing algorithms, it also make it possible to easily model and implement newer network services such as service chaining. It is a possible future direction in this study to investigate modelling various requirements in networking that are not available at the moment. Other possible future directions of this study might include optimising application modelling in terms of number of decision variables. Currently no search strategies are implemented in the applications and the interface. One possible enhancement to the current version would be to research and determine the optimised search strategies in SCOR. The other direction of the research for practical purposes might include the choice of appropriate solver for various QoS routing and TE applications. \\

\bibliographystyle{IEEEtran}
\bibliography{IEEEabrv,references} 


\end{document}